\documentclass[12pt]{iopart} 

\usepackage{epsfig} 
\usepackage{iopams}

\newcommand{\dt}{\Delta t}
\newcommand{\tw}{t_\mathrm{w}} 
\newcommand{\eq}{\mathrm{eq}}
\newcommand{\ceq}{\varepsilon}

\newcommand{\cphi}{\varphi} 
\newcommand{\Erfc}{\Phi} 

\newcommand{\bra}[1]{\langle #1 |}
\newcommand{\ket}[1]{| #1 \rangle}
\newcommand{\braket}[2]{\langle #1 | #2 \rangle} 

\newcommand{\ivec}{\boldsymbol{i}} 
\newcommand{\jvec}{\boldsymbol{j}} 
\newcommand{\kvec}{\boldsymbol{k}} 
\newcommand{\lvec}{\boldsymbol{l}} 

\newcommand{\bsig}{\boldsymbol{\sigma}}
\newcommand{\bn}{\boldsymbol{n}} 

\newcommand{\n}{\hat{n}} 
\newcommand{\G}[1]{G_{#1}^{(2)}} 
\newcommand{\Wc}{W^\mathrm{c}} 
\newcommand{\Wa}{W^\mathrm{a}} 
\newcommand{\Ws}{W^\sigma}
\newcommand{\Wfa}{W^\mathrm{FA}} 

\newcommand{\Vs}{V^{(\mathrm{s})}}
\newcommand{\Va}{V^{(\mathrm{a})}}

\newcommand{\rs}{R^{(\mathrm{s})}}
\newcommand{\ra}{R^{(\mathrm{a})}}

\newcommand{\rad}{R_{\mathrm{d}}^{(\mathrm{a})}} 
\newcommand{\xad}{\chi_{\mathrm{d}}^{(\mathrm{a})}} 
\newcommand{\rsd}{R_{\mathrm{d}}^{(\mathrm{s})}} 
\newcommand{\xsd}{\chi_{\mathrm{d}}^{(\mathrm{s})}} 

\newcommand{\rae}{R_{\mathrm{e}}^{(\mathrm{a})}} 
\newcommand{\rse}{R_{\mathrm{e}}^{(\mathrm{s})}} 
\newcommand{\xse}{\chi_{\mathrm{e}}^{(\mathrm{s})}} 

\newcommand{\Ds}{\Delta^{(\mathrm{s})}}
\newcommand{\Da}{\Delta^{(\mathrm{a})}}

\newcommand{\Des}{\Delta_{\mathrm{e}}^{(\mathrm{s})}}
\newcommand{\Dea}{\Delta_{\mathrm{e}}^{(\mathrm{a})}}

\newcommand{\cord}{C_\mathrm{d}} 
\newcommand{\resd}{R_\mathrm{d}} 
\newcommand{\core}{C_\mathrm{e}} 
\newcommand{\rese}{R_\mathrm{e}} 

\newcommand{\Xd}{X_\mathrm{d}} 
\newcommand{\Xe}{X_\mathrm{e}}

\begin{document} 
 
\title[Aging in coagulation-diffusion and the FA model]
{Aging in one-dimensional coagulation-diffusion processes and the Fredrickson-Andersen model} 

\author{Peter Mayer$^1$ and Peter Sollich$^2$}

\address{$^1$ Department of Chemistry, Columbia University, 
3000 Broadway, New York, NY 10027, USA}

\address{$^2$ King's College London, Department of Mathematics, Strand,
London WC2R 2LS, UK}

\ead{\mailto{pm2214@columbia.edu} and \mailto{peter.sollich@kcl.ac.uk}}
 
\begin{abstract} 
We analyse the aging dynamics of the one-dimensional Fredrickson-Andersen (FA) model 
in the nonequilibrium regime following a low temperature quench. Relaxation then 
effectively proceeds via diffusion limited pair coagulation (DLPC) of mobility excitations. 
By employing a familiar stochastic similarity transformation, we map exact results 
from the free fermion case of diffusion limited pair annihilation to DLPC. 
Crucially, we are able to adapt the mapping technique to 
averages involving multiple time quantities. This relies on knowledge of the explicit 
form of the evolution operators involved. Exact results are obtained for 
two-time correlation and response functions in the free fermion DLPC process. The 
corresponding long-time scaling forms apply to a wider class of DLPC processes, including the FA model. We are thus able to exactly characterise the violations of the 
fluctuation-dissipation theorem (FDT) in the aging regime of the FA model.
We find nontrivial scaling forms for the fluctuation-dissipation 
ratio (FDR) $X = X(\tw/t)$, but with a {\em negative} asymptotic value 
$X^\infty = -3\pi/(6\pi - 16) \approx -3.307$. 
While this prevents a thermodynamic interpretation in terms of an effective temperature, 
it is a direct consequence of probing FDT with observables that couple to 
activated dynamics. The existence of negative FDRs should therefore be a widespread feature 
in non mean-field systems. 
\end{abstract} 


\maketitle

\section*{Introduction} 

A generic feature of glass-formers is the rapid increase of their relaxation 
time with decreasing temperature. When quenched below the laboratory glass 
transition, where relaxation times exceed experimental timescales, these systems 
fall out of equilibrium. They change from equilibrated fluids to nonequilibrium 
amorphous solids \cite{EdiAngNag96}. The ``waiting time" $\tw$ elapsed since 
preparation of the state then sets a timescale for the subsequent relaxation: 
the system ages \cite{Young}. A full understanding of this nonequilibrium 
phenomenon remains a central theoretical challenge. 

Mean-field models of structural and spin glasses \cite{CugKur93,CugKur94}, in 
which relaxation times strictly diverge at the glass transition, have delivered 
important insights into the dynamics of aging. Most notably, these systems satisfy 
a generalized fluctuation-dissipation theorem (FDT) in their nonequilibrium, 
aging state. This is defined in terms of the two-time connected correlation 
function for a generic observable $A$, 
\begin{equation}
  C(t,\tw) = \langle A(t) A(\tw) \rangle - \langle A(t) \rangle \langle A(\tw) \rangle, 
  \label{equ:C2tdef}
\end{equation}
with $t \geq \tw$, and the corresponding two-time response function 
\begin{equation}
  R(t,\tw) = T \left. \frac{\delta \langle A(t) \rangle}{\delta h(\tw)} \right|_{h = 0}.  
  \label{equ:R2tdef}
\end{equation}
Here $h$ denotes the thermodynamically conjugate field to the observable $A$ so that 
the perturbation of the Hamiltonian is $\delta \mathcal{H} = - h A$. Note that we have absorbed the 
temperature $T$ in the definition of the response. The associated generalized FDT is 
then 
\begin{equation}
  R(t,\tw) = X(t,\tw) \frac{\partial}{\partial \tw} C(t,\tw), 
  \label{equ:gFDT}
\end{equation}
with $X(t,\tw)$ the fluctuation-dissipation ratio (FDR). At equilibrium, correlation 
and response functions are time translation invariant, depending only on $\dt = t - \tw$, 
and equilibrium FDT imposes that $X = 1$. This is no longer true in nonequilibrium 
systems. But the definition of an FDR through Eq.~(\ref{equ:gFDT}) becomes nontrivial 
for aging systems: in the mean-field spin glass models \cite{CugKur93,CugKur94} its 
dependence on both time arguments is only through the correlation function $X(t,\tw) 
\sim X(C(t,\tw))$ at large times. This led to the introduction of timescale dependent 
effective temperatures~\cite{CugKurPel97}, and a possible thermodynamic interpretation 
of aging~\cite{CugKur93,CugKur94,FraMezParPel98}.

In many systems of physical interest, however, such as liquids 
quenched below the glass transition or domain growth in disordered 
magnets \cite{FisHus86}, the dynamics is not of mean-field type. Crucial new features are 
activated processes and spatial heterogeneity~\cite{Sillescu99,Ediger00,Glotzer00,Andersen05}. 
Some experiments and simulations~\cite{CriRit03} have nonetheless seemed
to detect a mean-field aging regime. On the other hand, theoretical studies have found ill-defined 
FDRs~\cite{CriRitRocSel00,VioTalTar03}, non-monotonic response 
functions~\cite{Nicodemi99,GarNew00,DepSti05,Krzakala05}, observable 
dependence~\cite{FieSol02,MayBerGarSol03}, nontrivial FDRs without 
thermodynamic transitions~\cite{BuhGar02,JacBerGar06} and a subtle interplay 
between growing dynamical correlation lengthscales and FDT 
violations~\cite{BarBer01,CasChaCugKen02}. Also in experiments deviations from mean-field expectations have been observed, with for example anomalously large FDT violations associated with intermittent 
dynamics~\cite{BelCilLar01}.  It thus remains an important task to delineate 
when the mean-field concept of an FDR-related effective temperature 
remains viable. 

In this paper, we focus on the one-dimensional Fredrickson-Andersen (FA) 
model \cite{FreAnd84,RitSol03}. Its relaxation time follows an Arrhenius law 
at low temperatures, and it is the simplest kinetically constrained 
model for glassy systems which displays dynamical 
heterogeneity \cite{GarCha02,GarCha03,JunGarCha04,TonWyaBerBirBou05}. 
We use it to study systematically the impact of activated,
and in this sense strongly non mean-field, dynamics on FDRs and
associated effective temperatures. In addition, the FA model dynamics 
become critical at low temperatures 
\cite{WhiBerGar04,WhiBerGar05,JacMaySol06}, with a diverging dynamical 
lengthscale. Our work is also of relevance, therefore, to the study of FDRs 
in nonequilibrium critical dynamics. Here, FDRs play the role of
universal amplitude ratios, which makes them important markers for
distinguishing dynamic universality classes \cite{CalGam05}.

A first numerical study of FDT violation in the one-dimensional FA model 
appeared in \cite{CriRitRocSel00} and suggested ill defined FDRs. 
The authors used disconnected correlations, however, which compromises their 
results since FDT would not be recovered in equilibrium. More recent numerical 
studies \cite{BuhGar02,Buh03} based on the correct, connected correlations 
indicate a very different picture: {\em no} detectable violation of the 
equilibrium FDT is observed in spite of the nonequilibrium aging dynamics. We 
show below that this is a consequence of the specific scaling in the quasi-equilibrium 
regime \cite{MaySol05}, which acts to obscure genuine aging contributions. These contributions are nevertheless present as we will show, leading to 
FDT violations with well-defined FDRs in the activated aging regime of 
the FA model. We take advantage of observable 
dependence of the FDR in order to gain more direct access to FDT 
violations, by considering global rather than local observables. Our discussion also elucidates the physical origin of negative 
dynamical response functions, and predicts the generic existence of negative 
FDRs for observables directly coupled to activated processes. A partial summary of our 
results can be found in \cite{MayLeoBerGarSol06}. 

Our analysis of aging in the FA model is based on a close connection with 
diffusion limited pair coagulation (DLPC) processes. The research field 
of diffusion limited reaction processes is well developed \cite{Kampen81} 
and offers powerful analytical tools \cite{HenOrlSan97,TauHowLee05}. 
These include, in particular, stochastic similarity 
transformations \cite{HenOrlSch95,KrePfaWehHin95}. We use the latter to 
map exact results for the free fermion case of diffusion limited pair 
annihilation (DLPA) to DLPC. In particular, we show how the mapping technique 
can be adapted to averages involving multiple time quantities. This relies on 
knowledge of the explicit form of the evolution operators involved. We derive 
simple and explicit propagators for certain observables in DLPA and DLPC. 
In the free fermion DLPC process this yields new and exact results, for instance, 
the two-time particle correlation functions. Our propagators are of generic 
value for the analysis of DLPA and DLPC, however, and could serve in further 
studies of these processes. 

The paper is organized as follows: in Sec.~\ref{sec:basics} we recall the 
definition of the FA model and introduce the matrix formalism used throughout 
this manuscript. A discussion of the dynamics in the different stages of 
relaxation is given, and we establish the connection to DLPC. The mapping 
between DLPC and DLPA is then stated in Sec.~\ref{sec:map}. We introduce suitable 
sets of observables, which are related in a simple manner by the mapping, 
and give expressions for their dynamics in the free fermion processes. 
Arguments are presented that establish the broader validity of the predictions for the long-time scaling 
behaviour. The generalisation of the mapping procedure to multiple time 
averages and propagators is presented in Sec.~\ref{sec:2t}. Using the 
DLPC propagator, expressions for two-time correlation and response functions 
are derived in Sec.~\ref{sec:cor2t} and Sec.~\ref{sec:res2t}. Response 
functions in the FA model comprise two contributions representing different 
physical effects and correspondingly Sec.~\ref{sec:res2t} is split into two 
subsections. Readers more interested in results than in their derivation 
may skip Sec.~\ref{sec:map} and Sec.~\ref{sec:2t} on first reading, except 
for the beginning of Sec.~\ref{sec:res2t} where our decomposition of 
response functions is introduced. 
Results for FDT violation in the activated aging regime of 
the FA model are then presented in Sec.~\ref{sec:local_fdt} and 
Sec.~\ref{sec:global_fdt} and deal with local and global observables, 
respectively. Each of these sections comprises three subsections, discussing 
separately the dynamics of correlations, response functions and the resulting 
FDRs. We conclude with a summary and discussion of our results in 
Sec.~\ref{sec:summary}. Two short appendices contain useful mathematical 
and technical material.

\section{Matrix Formalism}
\label{sec:basics} 

In this section we summarize the background for our subsequent analysis 
of the aging dynamics in the FA model. After recalling the definition of 
the FA model we review the standard operator formalism widely used in 
the literature for studying reaction diffusion problems. We then translate 
the FA model into this representation and discuss its effective dynamics 
in the different stages of relaxation after a quench.

The one-dimensional FA model \cite{FreAnd84} is defined on a linear lattice of size $N$ 
subject to periodic boundary conditions. Assigned to each site $i$ is 
a binary variable $n_i \in \{0,1\}$, with $n_i=1$ representing a mobile 
fluid region and $n_i=0$ an immobile one. The state of the system is
characterised by $\bn = (n_1,n_2, \ldots, n_N)$. The probability $p(\bn,t)$ 
of being in state $\bn$ at time $t$ evolves according to the generic master 
equation 
\begin{equation}
  \partial_t \, p(\bn,t) = \sum_{\bn'} \left[ w(\bn' \to \bn) p(\bn',t) 
  - w(\bn \to \bn') p(\bn,t) \right], 
  \label{equ:master}
\end{equation}
where $\partial_t$ denotes the time derivative and $w(\bn \to \bn')$ the 
rate for transitions from state $\bn$ to $\bn'$. In the FA model these 
transition rates are given by 
\begin{equation}
  w(\bn \to \bn') = \sum_i f_i(\bn) w_i(\bn) \delta_{n_i,1-n_i'} \prod_{j \neq i} \delta_{n_j,n_j'}. 
  \label{equ:wFA} 
\end{equation}
The Kronecker deltas, $\delta_{n,n'} = 1$ if $n=n'$ and zero otherwise, 
express that transitions occur only between states $\bn, \bn'$ differing 
by a single flip $n_i \to 1-n_i$. Under Glauber dynamics at temperature 
$T$ the (unconstrained) flip-rates are 
\begin{equation}
	w_i(\bn) = \ceq (1-n_i) + (1-\ceq) n_i, 
	\label{equ:wiFA} 
\end{equation}
and obey detailed balance with respect to the noninteracting Hamiltonian 
$\mathcal{H} = \sum_i n_i$. Here $\ceq = \langle n_i \rangle = 1/(1+\rme^{1/T})$ 
is the equilibrium density of mobility excitations; at low
temperatures this is small and so the excitations can also be thought
of as defects. The key ingredient of the 
FA model is the kinetic constraint 
\begin{equation} 
  f_i(\bn) = n_{i-1} + n_{i+1}. 
  \label{equ:fi} 
\end{equation} 
The dynamics on each site $i$ require facilitation by an adjacent mobility 
excitation $n_{i-1} = 1$ or $n_{i+1}=1$. Since $f_i(\bn)$ is independent of 
$n_i$ it preserves detailed balance; the equilibrium state of the FA model 
is trivial. However, the kinetic constraint induces nontrivial relaxation 
dynamics with a rich phenomenology including dynamical heterogeneity 
\cite{GarCha02,GarCha03,JunGarCha04,TonWyaBerBirBou05,WhiBerGar04,WhiBerGar05,
JacMaySol06,MerGarCha05,JacGarCha06,Garrahanetal07}. 

The discussion throughout this manuscript is based on a standard operator 
for\-ma\-lism. In the following we briefly 
summarize the main ideas; more details may be found in \cite{HenOrlSan97,KrePfaWehHin95}. One associates 
with each site $i$ two orthonormal states $\ket{0}$ and $\ket{1}$, or $\ket{n_i}$ 
for short, representing the mobility $n_i$. Then an orthonormal basis for 
the configuration space of the entire system is 
\begin{equation}
  \ket{\bn} = \bigotimes_i \ket{n_i} 
  \quad \mbox{with} \quad 
  \braket{\bn}{\bn'} = \prod_i \delta_{n_i,n_i'}. 
  \label{equ:basis} 
\end{equation}
Probabilistic states $\ket{P(t)}$ are defined by 
\begin{equation}
  \ket{P(t)} = \sum_{\bn} p(\bn,t) \ket{\bn}, 
  \label{equ:ketPdef}
\end{equation}
from which the probability for any particular configuration $\bn$ is extracted 
via $p(\bn,t) = \braket{\bn}{P(t)}$. One now multiplies the master equation 
Eq.~(\ref{equ:master}) by $\ket{\bn}$ and takes a sum over $\bn$. The result 
may be cast in the form
\begin{equation}
  \partial_t \, \ket{P(t)} = W \ket{P(t)}, 
  \label{equ:masterW}
\end{equation}
with $W$ the master operator 
\begin{equation}
  W = \sum_{\bn,\bn'} w(\bn \to \bn') \left( \ket{\bn'} \bra{\bn} - \ket{\bn} \bra{\bn} \right). 
  \label{equ:W}
\end{equation}
Conservation of probability is compactly expressed by $\braket{e}{P(t)} = 1$ 
if one introduces 
\begin{equation}
  \bra{e} = \sum_{\bn} \bra{\bn} = (\bra{0} + \bra{1})^{\otimes N}. 
  \label{equ:brae} 
\end{equation}
This is the bra ground state of $W$ because $\bra{e} W = 0$ from 
Eq.~(\ref{equ:masterW}). The corresponding ket ground state is by 
definition the equilibrium state $\ket{P_{\eq}}$ so that $W$ is generally 
a non-hermitian operator. The formal solution of Eq.~(\ref{equ:masterW}) 
for initial state $\ket{P(0)}$ is 
\begin{equation}
  \ket{P(t)} = \rme^{W t} \ket{P(0)}, 
  \label{equ:evolution} 
\end{equation}
with $\rme^{W t}$ the evolution operator. To calculate the expectation 
value of an observable $A(\bn)$ one introduces the associated operator 
$A = \sum_{\bn} A(\bn) \ket{\bn}\bra{\bn}$ in terms of which 
\begin{equation}
  \langle A(t) \rangle = \bra{e} A \ket{P(t)}.   
  \label{equ:average}
\end{equation}
It is straightforward to verify that indeed $\bra{e} A \ket{P(t)} = \sum_{\bn} 
A(\bn) p(\bn,t)$. This completes our recap of the operator formalism. 

The FA master operator $\Wfa$ is obtained by substituting the transition rates 
Eq.~(\ref{equ:wFA}) into Eq.~(\ref{equ:W}). To state it explicitly it
is useful to introduce the number operator 
\begin{equation}
  \n_i = \sum_{\bn} n_i \ket{\bn}\bra{\bn} = (\ket{1}\bra{1})_i. 
  \label{equ:ni} 
\end{equation} 
The notation $(X)_i$ indicates that the operator $X$ lives in the subspace of 
site $i$, that is, $(X)_i = 1^{\otimes (i-1)} \otimes X \otimes 1^{\otimes (N-i)}$ 
with $1 = \ket{0}\bra{0} + \ket{1}\bra{1}$ the diagonal operator; we
will not distinguish the identity operator $1$ from the number $1$ in our notation
as the meaning is always clear from the context. We also introduce the flip operator 
\begin{equation} 
	F_i  = (\ket{1}\bra{0} + \ket{0}\bra{1})_i,  
	\label{equ:Fi}
\end{equation} 
which maps states $\ket{\bn}$ onto $F_i \ket{\bn} = \ket{n_1, \ldots, n_{i-1}, 1-n_i, 
n_{i+1}, \ldots, n_N}$. It is important to note that $F_i$ and $\n_i$ do not commute. 
But of course $[F_i,\n_j]=0$ if $i \neq j$ since both operators act locally. Other 
obvious and useful properties are $\n_i^2 = \n_i$ and $F_i^2 = 1$. The FA master 
operator now reads 
\begin{equation}
  \Wfa = \sum_i (F_i - 1) f_i w_i, 
  \label{equ:WFAaux} 
\end{equation}
where $f_i = \n_{i-1} + \n_{i+1}$ and $w_i = \ceq (1-\n_i) + (1-\ceq) \n_i$. 
As will become clear immediately it is useful to substitute the explicit form 
of $f_i$ and, through a shift of the summation variable, rewrite the master 
operator as  
\begin{equation}
  \Wfa = \sum_i \Wfa_i, 
  \qquad  
  \Wfa_i = (F_{i+1} - 1) \n_i w_{i+1} + (F_i - 1) \n_{i+1} w_i. 
  \label{equ:WFA} 
\end{equation}
The key feature of Eq.~(\ref{equ:WFA}) is that $\Wfa_i$ operates only in the subspace 
$\ket{n_i,n_{i+1}}$. It therefore represents a {\em pair reaction process}. In the 
analysis of pair reaction processes it is conventional to choose an explicit 
representation for the basis $\ket{0} = {1 \choose 0}$ and $\ket{1} = {0 \choose 1}$. 
All operators then have appealingly simple matrix representations. From Eqs.~(\ref{equ:ni}) 
and (\ref{equ:Fi}) one has $\n_i = {{0 \,\,\, 0} \choose {0 \,\,\, 1}}_i$ and 
$F_i = {{0 \,\,\, 1} \choose {1 \,\,\, 0}}_i$. In particular, the FA master operator 
in the $\ket{n_i,n_{i+1}}$ subspace becomes 
\begin{equation}
  \Wfa_i = 
  \left( 
  \begin{array}{cccc}
    0 & 0      & 0      & 0        \\ 
    0 & -\beta & 0      & \gamma   \\ 
    0 & 0      & -\beta & \gamma   \\ 
    0 &  \beta &  \beta & -2\gamma  
  \end{array}
  \right)_{i,i+1},
  \label{equ:WFA4x4} 
\end{equation}
with $\beta = \ceq$ and $\gamma = 1-\ceq$. The rows and columns of this matrix 
correspond to the configurations $0 0$, $0 1$, $1 0$ and $1 1$ from top to bottom and 
left to right, respectively. Consequently $\beta$ is the branching
rate, i.e.\ the rate for processes 
$01 \to 11$ and $10 \to 11$, while $\gamma$ is the rate for coagulations $11 \to 01$ and 
$11 \to 10$. 

We emphasize that the transitions $101 \to 111$ or $111 \to 101$ are accounted for correctly within this pair reaction description. 
In the case $101 \to 111$, for instance, one has $f_i(\bn) = 2$ and $w_i(\bn) = \ceq$ 
according to Eqs.~(\ref{equ:wiFA}) and (\ref{equ:fi}), which amounts to a transition rate of 
$2 \ceq$. Equivalently, from Eqs.~(\ref{equ:WFA}) and (\ref{equ:WFA4x4}) the process 
$101 \to 111$ is achieved by either branching to the right $\underline{10}1 \to \underline{11}1$ 
or to the left $1\underline{01} \to 1\underline{11}$ so that its total rate is again 
$\ceq + \ceq = 2 \ceq$ as required. The pair reaction picture is of course exact. 

We note that a modified version of the FA model with kinetic constraint 
$f_i(\bn) = n_{i-1} + n_{i+1} - n_{i-1} n_{i+1}$ has often
been used in the literature since it was first proposed \cite{GraPicGra93}. It
has certain advantages for simulations because $f_i\in\{0,1\}$ is just
a binary variable indicating whether the constraint of having at least
one mobile neighbour is satisfied. However, for analytical work the
presence of the quadratic term $n_{i-1}n_{i+1}$ is awkward and
prevents in particular a pair reaction representation like the one above. 
Nevertheless the scaling behaviour in the modified FA model is expected 
to match with the original version \cite{FreAnd84} discussed here.

\subsection{Effective Dynamics} 
\label{sec:effective} 

The branching and coagulation process Eq.~(\ref{equ:WFA4x4}) gives rise to the 
remarkably rich dynamics of the FA model. 
Here we are interested in the evolution after a quench from a random uncorrelated initial 
state (equilibrium at $T = \infty$) to some low temperature $T \ll 1$ where $\ceq \sim \rme^{-1/T} \ll 1$. 
It is well known that the dynamics then evolves in distinct stages with 
separated time scales \cite{RitSol03}. 

On the $\mathcal{O}(1)$ time scale activated states relax into blocked 
configurations. That is, the dynamics is controlled by coagulation processes 
$11 \to 01$ or $10$ which proceed with rate $\gamma \sim 1$ until no further 
activated states $11$ remain. This process has been analyzed in 
\cite{FolRit96,SchTri97}. The density of excitations drops from its initial 
value of $c_0=\frac{1}{2}$ to $c_0 \exp(-c_0)$. 
Branching events like $10 \to 11$ only become relevant for times $t = 
\mathcal{O}(1/\ceq)$ since their rate is $\beta = \ceq \ll 1$. 
Such processes increase the energy $\mathcal{H} = \sum_i n_i$ and 
thus require thermal activation. The lifetime of the resulting 
activated state is $\mathcal{O}(1)$ because the excitations still coagulate 
with rate $\gamma \sim 1$, either back into the original state $10$ 
or into $01$, each with equal probability. Due to this separation 
of time scales the probability to find an activated state at any given 
moment $t$ is $\mathcal{O}(\ceq)$. At low temperatures $\ceq 
\ll 1$ branching events therefore effectively reduce to diffusion 
steps $01 \leftrightharpoons 10$, the rate of which is $d = \beta/2 
= \ceq/2$. The FA dynamics on the $\mathcal{O}(1/\ceq)$ 
time scale is effectively that of diffusion and coagulation 
\cite{WhiBerGar04,WhiBerGar05,JacMaySol06}, 
\begin{equation}
  \Wc_i = 
  \left( 
  \begin{array}{cccc}
    0 & 0  & 0  & 0        \\ 
    0 & -d & d  & \gamma   \\ 
    0 & d  & -d & \gamma   \\ 
    0 & 0  & 0  & -2\gamma  
  \end{array}
  \right)_{i,i+1}.
  \label{equ:WFAeff} 
\end{equation}
This statement becomes exact in the limit $\ceq \to 0$ at fixed 
scaled time $\tilde{t} = \ceq t$. In other words, the evolution 
operators satisfy $\exp(\Wfa \tilde{t}/\ceq) \sim 
\exp(\Wc \tilde{t}/\ceq)$. In units of scaled time $\tilde{t}$ 
the effective master operator $\tilde{W}^\mathrm{c} = \ceq^{-1} \Wc$ 
has diffusion rate $\tilde{d} = 1/2$ while the coagulation rate 
$\tilde{\gamma} \sim 1/\ceq \to \infty$ becomes formally infinite. 
(The process nevertheless remains well-defined because it is diffusion 
limited.)

At small but finite $\ceq \ll 1$ a third dynamical regime appears 
for times $t = \mathcal{O}(1/\ceq^2)$. Double-branching events 
$100 \to 110 \to 111$, which have rate $\mathcal{O}(\ceq^2)$, 
can then no longer be neglected. Such triples relax with 
probability $1/2$ into $101$, so that two stable mobility excitations emerge. This 
process continuously generates new mobility excitations and thus 
counteracts the diffusion and coagulation dynamics discussed above. 
On a time scale of $\mathcal{O}(1/\ceq^2)$ or larger a crossover must
therefore set in 
that eventually drives the FA model into equilibrium. A simple argument 
shows \cite{RitSol03} that equilibrium is attained only for times of $\mathcal{O}(1/\ceq^3)$.

In our analysis of aging in the FA model we focus on the nonequilibrium 
regime of times $1 \ll t \ll 1/\ceq^2$ or more precisely scaled 
times $\tilde{t} = \ceq t = \mathcal{O}(1)$. In units of $\tilde{t}$ 
the evolution is then governed by the diffusion and coagulation 
process Eq.~(\ref{equ:WFAeff}) with rates $\tilde{d} = 1/2$ and 
$\tilde{\gamma} \sim 1/\ceq \gg 1$. We will concentrate in particular 
on the scaling that emerges for large $\tilde{t}$. At finite temperature 
$\ceq > 0$ this scaling refers to times $\tilde{t} \gg 1$ but 
still $\tilde{t} \ll 1/\ceq$ since Eq.~(\ref{equ:WFAeff}) does not 
account for the onset of branching processes at $t = 
\mathcal{O}(1/\ceq^2)$. In other words, whenever we talk 
about asymptotic behaviour for $\tilde{t} \to \infty$ it is 
understood that the zero-temperature limit $\ceq \to 0$ is 
taken first. 

We will later want to analyze response functions resulting from local 
perturbations $\delta\mathcal{H} = - \sum_i h_i n_i$. The impact of the 
latter on the effective dynamics Eq.~(\ref{equ:WFAeff}) is easily derived. 
In the Glauber rates Eq.~(\ref{equ:wiFA}), which are defined through 
$w_i(\bn) = \left[1 + \exp(\Delta_i \mathcal{H}/T) \right]^{-1}$ with 
$\Delta_i \mathcal{H}$ the change in energy caused by the flip, $\ceq$ 
is then replaced by the local equilibrium excitation density 
\begin{equation}
  \varepsilon_i = \frac{1}{1+\rme^{(1-h_i)/T}} \approx \ceq (1+h_i/T), 
  \label{equ:c_i}
\end{equation}
to linear order in $h_i$. Consequently the branching rates from sites 
$i-1$ and $i+1$ to site $i$ in the FA master operator Eq.~(\ref{equ:WFA4x4}) 
become $\beta = \varepsilon_i$; corresponding changes in the coagulation rates 
$\gamma = 1 + \mathcal{O}(\ceq)$ are negligible at low temperature. 
In the effective dynamics Eq.~(\ref{equ:WFAeff}) 
on the $\mathcal{O}(1/\ceq)$ time scale we then have
$d = \beta/2 = \varepsilon_i/2$  for diffusion from sites 
$i-1$ and $i+1$ to site $i$ or, in scaled time $\tilde{t}$, 
$\tilde{d} \approx 1/2 + h_i/(2T)$ and $\tilde{\gamma} = \mathcal{O}(1/\ceq)$.
In a linear response calculation on the $\tilde{t} = \mathcal{O}(1)$ 
time scale and at low temperature $\ceq \ll 1$, a field $h_i$ 
thus enhances the rates for diffusion towards site $i$. Physically, this effect arises because diffusion in the FA model is an activated process.

\section{A Useful Mapping}
\label{sec:map}

The effective DLPC dynamics of the FA model on the $\mathcal{O}(1/\ceq)$ 
time scale can be mapped onto a DLPA 
process. The latter admits an exact solution only for a special 
set of rates. We argue below that the scaling of certain observables, 
which includes in particular the domain size distribution, is robust 
under this modification of the rates. This will allow us to obtain various exact scaling 
results for the low temperature dynamics. We always focus on the
$\mathcal{O}(1/\ceq)$ time scale; to
simplify the notation we will drop all tildes from now on, effectively using
$1/\ceq$ as the unit of time.

The equivalence of entire classes of diffusion limited reaction processes 
has been established in \cite{HenOrlSch95}. 
The effective DLPC dynamics, Eq.~(\ref{equ:WFAeff}),
of the FA model in particular is equivalent to the process 
\begin{equation}
  \Wa_i = \left( \begin{array}{rrrc} 
    0 &  0 &  0 & \alpha  \\ 
    0 & -d & d  & 0       \\ 
    0 & d  & -d & 0       \\ 
    0 &  0 &  0 & -\alpha 
  \end{array} \right)_{i,i+1}, 
  \label{equ:Wa}
\end{equation}
which describes diffusion limited pair annihilation. Here $d$ is 
the rate for diffusion $10 \leftrightharpoons 01$, while $\alpha$ is the rate for 
pair annihilation $11 \to 00$. This result is established through a local 
stochastic similarity transformation with $b= \ket{0}\bra{0} + 2 \ket{1}\bra{1} 
- \ket{0}\bra{1}$, or in matrix representation \cite{KrePfaWehHin95}, 
\begin{equation}
  b = \left( \begin{array}{rr} 1 & -1 \\ 0 & 2 \end{array} \right) 
  \quad \mbox{with} \quad 
  b^{-1} = \left( \begin{array}{rr} 1 & \hspace{1ex} \frac{1}{2} \\ 0 & \frac{1}{2} 
  \end{array} \right). 
  \label{equ:b}
\end{equation}
In terms of 
\begin{equation}
  B = b^{\otimes N} \quad \mbox{and} \quad  B^{-1} = \left( b^{-1} \right)^{\otimes N}, 
  \label{equ:B}
\end{equation}
one has 
\begin{equation}
  \Wc = B \Wa B^{-1}, 
  \label{equ:map}
\end{equation}
if $\Wc$ and $\Wa$ share the same diffusion rate $d$ but $\alpha = 2\gamma$. 
This is readily proved: since the transformation is local and because 
$W^{\mathrm{c},\mathrm{a}} = \sum_i W_i^{\mathrm{c},\mathrm{a}}$ are pair 
reaction processes it is sufficient to consider the local master operators 
$W_i^\mathrm{a}$ and $W_i^\mathrm{c}$ in the $\ket{n_i,n_{i+1}}$ subspace. 
A simple matrix multiplication then shows that 
$W_i^\mathrm{c} = (b \otimes b) W_i^\mathrm{a} (b^{-1} \otimes b^{-1})$. 

We note as an aside that there is in fact a direct mapping \cite{JacMaySol06} that
establishes the equivalence  
of the FA model and the DLPA process Eq.~(\ref{equ:Wa}) on the 
$\mathcal{O}(1/\ceq)$ time scale, without having to proceed via the effective
DLPC dynamics. The FA master operator 
Eq.~(\ref{equ:WFA4x4}) exactly maps onto DLPA, Eq.~(\ref{equ:Wa}), 
but with an additional pair creation process $00 \to 11$ of 
rate $\mathcal{O}(\ceq^2)$. The latter is irrelevant for the 
dynamics on the $\mathcal{O}(1/\ceq)$ time scale so that we are 
once again left with the DLPA process Eq.~(\ref{equ:Wa}).

Let us now consider how observables transform under the mapping $B$. 
The dynamics of a generic observable $A$ in DLPC is given by 
\begin{equation}
  \langle A(t) \rangle_c = \bra{e} A \, \rme^{\Wc t} \ket{P(0)}. 
  \label{equ:Acdef} 
\end{equation}
By inserting identity operators $B B^{-1}$ and via Eq.~(\ref{equ:map}) 
this becomes 
\begin{equation}
  \langle A(t) \rangle_c = \bra{e} B \tilde{A} \, \rme^{\Wa t} B^{-1} \ket{P(0)}, 
  \label{equ:Aa1} 
\end{equation}
where $\tilde{A} = B^{-1} A B$. Note that $\bra{e} B = 
[ (\bra{0} + \bra{1}) \, b ]^{\otimes N} = \bra{e}$ according 
to Eqs.~(\ref{equ:brae}) and (\ref{equ:b}); likewise 
$\bra{e} B^{-1} = \bra{e}$. In order to eliminate $B^{-1}$ in 
Eq.~(\ref{equ:Aa1}) an initial state $\ket{P(0)}$ must be specified. 
Consider the example of an uncorrelated and homogeneous initial 
state with particle density $0 < \rho \leq 1$, 
\begin{equation}
  \ket{P(0)} = \ket{\rho} = \left[ (1-\rho) \ket{0} + \rho \ket{1} \right]^{\otimes N}. 
  \label{equ:ketrho} 
\end{equation}
Then $B^{-1} \ket{\rho} = \left[ (1-\rho) \, b^{-1} \ket{0} + 
\rho \, b^{-1} \ket{1} \right]^{\otimes N} = \left[ (1-\frac{1}{2}\rho) \ket{0} + 
\frac{1}{2}\rho \ket{1} \right]^{\otimes N} = \ket{\frac{1}{2} \rho}$ and 
\begin{equation}
  \langle A(t) \rangle_c 
  = \bra{e} A \, \rme^{\Wc t} \ket{\rho} 
  = \bra{e} \tilde{A} \, \rme^{\Wa t} \ket{{\textstyle \frac{1}{2}}\rho} 
  = \langle \tilde{A}(t) \rangle_a. 
  \label{equ:Aca} 
\end{equation}
This equation states that the dynamics of an observable $A$ in 
DLPC with initial density $\rho$ is identical to that of $\tilde{A}$ in 
DLPA for initial density $\rho/2$. One has to bear in mind, however,
that $\tilde{A}$ is generally a non-diagonal operator and so cannot be
interpreted as a standard physical observable. 
Setting $A = \n$, for instance, yields 
\begin{equation} 
  \tilde{n} = b^{-1} \n \, b = \ket{0}\bra{1} + \ket{1} \bra{1}. 
  \label{equ:ntilde} 
\end{equation}
The usefulness of Eq.~(\ref{equ:Aca}) is preserved only by the fact that $\tilde{A}$ 
is projected onto $\bra{e}$. For the number operator $(\bra{0} + \bra{1}) \, 
\tilde{n} = 2 \bra{1} = 2 (\bra{0} + \bra{1}) \hat{n}$ as one easily verifies. 
More generally this implies \cite{KrePfaWehHin95}
\begin{equation}
  \bra{e} \n_{i_1} \n_{i_2} \ldots \n_{i_m} \, \rme^{\Wc t} \ket{\rho} = 
  2^m \bra{e} \n_{i_1} \n_{i_2} \ldots \n_{i_m} \, \rme^{\Wa t} \ket{{\textstyle \frac{1}{2}}\rho}, 
  \label{equ:mapn} 
\end{equation}
for ordered (and hence distinct) $i_1 < i_2 < \ldots < i_m$. This relates 
equal-time particle correlations of arbitrary order in DLPC and DLPA
in a simple manner.
A more convenient set of observables for our purposes will be ($i_1 < i_2$) 
\begin{equation}
  E_{i_1,i_2} = \prod_{k=i_1}^{i_2-1} (1-\n_k) 
  \quad \mbox{and} \quad 
  P_{i_1,i_2} = \prod_{k=i_1}^{i_2-1} (1-2\n_k), 
  \label{equ:EPdef} 
\end{equation}
which we refer to as empty and parity interval operators, respectively \cite{BenAvraham98,MasAvr01}. 
This is for obvious reasons: $E_{i_1,i_2} \ket{P(t)} = \ket{P(t)}$ if the interval 
$\{i_1, i_1+1, \ldots, i_2-1\}$ contains no particles, i.e.\ is empty, and 
$E_{i_1,i_2} \ket{P(t)} = 0$ otherwise. Similarly $P_{i_1,i_2} \ket{P(t)} = 
(-1)^m \ket{P(t)}$ measures the parity of the number of particles $m$ contained 
in the same interval. From Eq.~(\ref{equ:Aca}) one has the transformation law 
\begin{equation}
  \bra{e} E_{i_1,i_2} \ldots E_{i_{2k-1},i_{2k}} \, \rme^{\Wc t} \ket{\rho} = 
  \bra{e} P_{i_1,i_2} \ldots P_{i_{2k-1},i_{2k}} \, \rme^{\Wa t} \ket{{\textstyle \frac{1}{2}}\rho}, 
  \label{equ:mapEP} 
\end{equation}
again assuming the $i_1 < i_2 < \ldots < i_{2k}$ are ordered. This expression 
is of course equivalent to Eq.~(\ref{equ:mapn}). But its physical meaning is 
more appealing: empty interval probabilities in DLPC directly map onto parity 
probabilities in DLPA if the initial state is adjusted appropriately. 

The mapping establishes that equivalent to the effective DLPC dynamics 
of the FA model, Eq.~(\ref{equ:WFAeff}), with rates $d = 1/2$ and 
$\gamma \sim 1/\ceq \gg 1$, is the DLPA process of Eq.~(\ref{equ:Wa}) 
with rates $d = 1/2$ and $\alpha = 2 \gamma \sim 2/\ceq \gg 1$. 
However, no exact solutions are known for either of 
these two processes. In DLPC the only exception is the process with identical 
diffusion and coagulation rates, $\gamma = d$, for which the master operator 
$\Wc$ admits a free fermion mapping \cite{HenOrlSch95}; obviously the same is true in 
the equivalent DLPA process with $\alpha = 2d$. In the following we 
analyze the dynamics of these particular free fermion DLPC and
DLPA processes in more detail. Thereafter we discuss to what extent 
the results depend on reaction rates. 

Exact results for reaction-diffusion problems are often related to 
integrable quantum spin chains \cite{HenOrlSan97}. Indeed, the most 
efficient way to generate explicit results for our free fermion 
DLPC and DLPA processes is to note that these are in turn equivalent 
to zero temperature dynamics of the Glauber Ising spin chain \cite{Glauber63}. 
%
%A number of exact results have been derived for free fermion DLPC and
%DLPA \cite{HenOrlSan97}. 
%The most efficient way to generate explicit expressions is to note that these 
%processes are in turn equivalent to the zero temperature dynamics of the 
%Glauber Ising spin chain \cite{Glauber63}. 
The latter is defined on a one-dimensional lattice $i = 1,2, \ldots, N$ of 
Ising spins $\sigma_i \in \{-1,+1\}$. The probability $p(\bsig,t)$ for 
being in state $\bsig = (\sigma_1, \sigma_2, \ldots, \sigma_N)$ at time 
$t$ is governed by a master equation of the form Eq.~(\ref{equ:master}) 
if one replaces $\bn$ by $\bsig$. Likewise the transition rates are of the 
form Eq.~(\ref{equ:wFA}) but without kinetic constraint, i.e.\ effectively $f_i(\bsig) = 1$. 
The spin flip rates in the Glauber Ising model at $T=0$ are $w_i(\bsig) = 
\frac{1}{2} [ 1 - \frac{1}{2} \sigma_i (\sigma_{i-1} + \sigma_{i+1})]$ 
and represent nearest neighbour ferromagnetic interactions. A connection 
to DLPA is established by considering the defect or domain wall observables 
$n_i = \frac{1}{2} ( 1 - \sigma_i \sigma_{i+1}) \in \{0,1\}$. Writing 
$\uparrow$ and $\downarrow$ for spin configurations $\sigma = +1$ and 
$\sigma = -1$, respectively, transitions $\uparrow \uparrow \downarrow 
\, \leftrightharpoons \, \uparrow \downarrow \downarrow$ then correspond 
to diffusion $0 1 \leftrightharpoons 1 0$ of the domain wall. The rate for 
this is $d = w_i(\bsig) = \frac{1}{2}$. Similarly $\uparrow \downarrow \uparrow 
\, \to \, \uparrow \uparrow \uparrow$ translates into annihilation $1 1 \to 0 0$ 
of domain walls and has rate $\alpha = w_i(\bsig) = 1$ (which satisfies 
the free fermion condition $\alpha = 2d$). The reverse process of pair 
creation $00 \to 11$ is not possible under zero temperature dynamics. 
Together, these arguments show that the spin dynamics in the Glauber
Ising chain correspond  
to a DLPA process of the domain walls \cite{Santos97}. We now exploit this connection 
to extract results for DLPA. It is convenient to introduce an operator 
formalism analogous to Eqs.~(\ref{equ:basis}) - (\ref{equ:average}) by 
simply replacing the $\ket{0}$, $\ket{1}$ basis with 
the local spin basis $\ket{\!\uparrow}$, $\ket{\!\downarrow}$. Further 
denote by $\sigma^z$ the spin operator $\sigma^z = \ket{\!\uparrow} 
\bra{\uparrow\!} - \ket{\!\downarrow} \bra{\downarrow\!}$ and 
by $\Ws = \sum_i \left( \sigma_i^x - 1 \right) w_i$ the master operator of the 
Glauber Ising chain; here $\sigma_i^x = (\ket{\!\uparrow} \bra{\downarrow\!} + 
\ket{\!\downarrow} \bra{\uparrow\!})_i$ plays the role of the flip operator 
Eq.~(\ref{equ:Fi}). Using that the domain wall operator in the spin basis is 
$\n_i = \frac{1}{2} \left( 1 - \sigma_i^z \sigma_{i+1}^z \right)$ the 
parity operator for the number of domain walls becomes 
\begin{equation}
  P_{i_1,i_2} = \prod_{k=i_1}^{i_2-1} (1-2\n_k) = \prod_{k=i_1}^{i_2-1} \sigma_k^z \sigma_{k+1}^z 
  = \sigma_{i_1}^z \sigma_{i_2}^z, 
  \label{equ:Ps} 
\end{equation}
since $\left(\sigma^z\right)^2 = 1$. This equation expresses the obvious 
fact that if $\sigma_{i_1}$ and $\sigma_{i_2}$ are aligned there must 
be an even number of sign-changes, i.e., domain walls, in the region 
between them. The fact that domain walls perform DLPA combined with 
Eq.~(\ref{equ:Ps}) supplies us with the result 
\begin{equation}
  \bra{e} P_{i_1,i_2} \ldots P_{i_{2k-1},i_{2k}} \, \rme^{\Wa t} \ket{{\textstyle \frac{1}{2}}\rho} = 
  \bra{e} \sigma_{i_1}^z \sigma_{i_2}^z \ldots \sigma_{i_{2k}}^z \, \rme^{\Ws t} \ket{P_\sigma(0)}, 
  \label{equ:Psmap} 
\end{equation}
for $i_1 < i_2 < \ldots < i_{2k}$ as usual. The dynamics of parity probabilities 
in the DLPA process with diffusion rate $d=1/2$ and annihilation rate $\alpha = 1$ 
are identical to spin correlations in the zero temperature dynamics of the Glauber 
Ising chain. The initial state of the spin system $\ket{P_\sigma(0)}$ corresponding 
to the uncorrelated and homogeneous initial state $\ket{\frac{1}{2} \rho}$ in DLPA 
is determined by setting $t=0$ in Eq.~(\ref{equ:Psmap}). Then $\bra{e} P_{i_1,i_2} 
\ldots P_{i_{2k-1},i_{2k}} \ket{\frac{1}{2} \rho} = \bra{e} P_{i_1,i_2} \ket{\frac{1}{2} \rho} 
\ldots \bra{e} P_{i_{2k-1},i_{2k}} \ket{\frac{1}{2} \rho}$ and $\bra{e} P_{i,j} 
\ket{\frac{1}{2} \rho} = (1-\rho)^{j-i}$ factorizes completely since 
$\ket{\frac{1}{2} \rho}$ is uncorrelated and thus~\footnote{
  Here the thermodynamic limit $N \to \infty$ was taken. In a finite system DLPA 
  conserves the parity $P_{1,N}$ of the total particle number. The sectors of 
  even and odd parity evolve independently and map onto Ising systems with periodic 
  and antiperiodic boundary conditions, respectively. In the $N \to \infty$ limit 
  the two sets of boundary conditions give identical results and so do
  not have to be distinguished.
}
\begin{equation}
  \bra{e} \sigma_{i_1}^z \sigma_{i_2}^z \ldots \sigma_{i_{2k}}^z \ket{P_\sigma(0)} 
  = \prod_{\lambda=1}^k (1-\rho)^{i_{2\lambda} - i_{2\lambda-1}}. 
\end{equation}
These are precisely the equilibrium correlations, $\ket{P_\sigma(0)} =
\ket{P_{\eq}(T_0)}$,  
for the Ising Hamiltonian $\mathcal{H} = - J \sum_i \sigma_i \sigma_{i+1}$ at temperature 
$T_0$ if  
\begin{equation}
  \tanh(J/T_0) = 1 - \rho.
  \label{equ:rhoT0} 
\end{equation}
Altogether, from Eqs.~(\ref{equ:mapEP}) and (\ref{equ:Psmap}), empty interval 
probabilities in the DLPC process with identical diffusion $d=1/2$ and coagulation 
$\gamma = 1/2$ rates are thus given by 
\begin{equation}
  \bra{e} E_{i_1,i_2} \ldots E_{i_{2k-1},i_{2k}} \, \rme^{\Wc t} \ket{\rho} = 
  \bra{e} \sigma_{i_1}^z \sigma_{i_2}^z \ldots \sigma_{i_{2k}}^z \, \rme^{\Ws t} \ket{P_{\eq}(T_0)}, 
  \label{equ:Esmap}
\end{equation}
where $i_1 < i_2 < \ldots < i_{2k}$. Equation (\ref{equ:Esmap}) is the central 
result of this section. It directly relates spin correlations in the Glauber Ising 
chain after a quench from temperature $T_0 > 0$ to $T=0$ to the dynamics of (multi) 
empty interval observables in DLPC for initial density $0 < \rho \leq 1$. The 
required spin correlations appearing on the right hand side of Eq.~(\ref{equ:Esmap}) 
were derived in \cite{MaySol04}. The simplest result is obtained for the initially 
filled state $\rho = 1$ which corresponds to $T_0 = \infty$ according to 
Eq.~(\ref{equ:rhoT0}). One then has \cite{MaySol04}
\begin{equation}
  \bra{e} E_{i_1,i_2} \ldots E_{i_{2k-1},i_{2k}} \, \rme^{\Wc t} \ket{1} = 
  \sum\limits_{\pi\in\mathcal{P}(k)} 
  (-1)^{\pi} \prod\limits_{\lambda=1}^k H_{i_{\pi(2\lambda)}-
  i_{\pi(2\lambda-1)}}(2t).  
  \label{equ:multiE}
\end{equation}
Here $\pi: \{1,2,\ldots, 2k\} \mapsto  \{1,2,\ldots, 2k\}$ denotes an index permutation, 
and $(-1)^\pi$ its sign; the sum runs over all permutations
in the set $\mathcal{P}(k)$ of ordered pairings 
of the numbers $\{1,2,\ldots, 2k \}$. In an ordered pairing the
entries are ordered within pairs, $\pi(1)<\pi(2)$, $\pi(3)<\pi(4)$, \ldots, 
$\pi(2k-1)<\pi(2k)$, and the pairs are arranged in order of their first entry so that
$\pi(1)<\pi(3)< \ldots < \pi(2k-1)$.
An expression for the function $H_n(t)$ 
is given in Eq.~(\ref{equ:Hndef}). We remark that Eq.~(\ref{equ:multiE}) 
also applies in the case $\rho < 1$, i.e.\ $T_0 < \infty$, if one replaces 
$H_n$ by the function $H_n''$ given in \cite{MaySol04}. In subsequent sections we will 
repeatedly make use of the result Eq.~(\ref{equ:multiE}) for the special cases 
$k=1,2$ where it reduces to 
\begin{eqnarray}
  \fl \bra{e} E_{i_1,i_2} \, \rme^{\Wc t} \ket{1} = H_{i_2-i_1}(2t),  
  \label{equ:E1t} \\
  \fl \bra{e} E_{i_1,i_2} E_{i_3,i_4} \, \rme^{\Wc t} \ket{1} = \left[ 
  H_{i_2-i_1} H_{i_4-i_3} - H_{i_3-i_1} H_{i_4-i_2} + 
  H_{i_4-i_1} H_{i_3-i_2} \right](2t). 
  \label{equ:E2t} 
\end{eqnarray}
Both expressions, like Eq.~(\ref{equ:multiE}), only apply for strictly ordered 
indices. In Eq.~(\ref{equ:E2t}) we have introduced the 
short hand notation $[ \, \cdot \, ](x)$ to indicate that all functions enclosed 
in the square brackets have the same argument $x$.

\subsection{Domain Size Distribution}
\label{sec:domain}

At this point we have the exact result Eq.~(\ref{equ:multiE}) for empty 
interval densities in the free fermion DLPC process, with rates $d = \gamma = 1/2$. 
However, we are actually interested in the effective FA process which
has the correct diffusion rate $d = 1/2$ but a much larger coagulation rate
$\gamma\sim 1/\ceq \gg 1$. In order to 
understand how the latter
affects the dynamics of DLPC it is instructive to consider the domain 
size distribution. Using spatial homogeneity we define the density of 
domains of size $k=j-i \geq 1$ as 
\begin{equation}
  D_{j-i}(t) = \bra{e} \n_i (1-\n_{i+1}) \cdots (1-\n_{j-1}) \n_j \, \rme^{\Wc t} \ket{1}. 
  \label{equ:Dkdef} 
\end{equation}
Equivalently $D_{j-i}(t) = \bra{e} \left( E_{i+1,j} - E_{i,j} - E_{i+1,j+1} 
+ E_{i,j+1} \right) \, \rme^{\Wc t} \ket{1}$ as follows by re\-writing 
$\n_i = 1 - ( 1 - \n_i )$ and similarly for $\n_j$. In the free fermion 
case these empty interval densities are given by Eq.~(\ref{equ:E1t}) so 
that $D_k(t) = [H_{k-1} - 2 H_k + H_{k+1}](2t)$. Applying the 
recursion Eq.~(\ref{equ:Hnrec}) turns this expression into 
\begin{equation}
  D_k(t) = \rme^{-2t} [I_{k-1} - I_{k+1}](2t), 
  \label{equ:DkI}
\end{equation}
where the $I_n(t)$ are modified Bessel functions, see Eq.~(\ref{equ:Inint}). 
This remarkably simple exact result is well known \cite{DerZei96}. Since one 
particle may be associated with each domain the exact particle concentration 
$c(t) = \langle \n_i(t) \rangle$ follows by summing this over $k$. The sum 
is telescopic and hence 
\begin{equation}
  c(t) = \sum_{k=1}^\infty D_k(t) = \rme^{-2t} [I_0 + I_1](2t). 
  \label{equ:cn}
\end{equation}
Obviously the results Eqs.~(\ref{equ:DkI}) and (\ref{equ:cn}) apply exactly 
only for the free fermion ($d = \gamma = 1/2$) DLPC process with
initial density $\rho = 1$.
However, it is known that both the concentration and the domain size
distribution have remarkably robust scalings \cite{ZhoBen95}. 
From Eq.~(\ref{equ:cn}) and the asymptotic expansion Eq.~(\ref{equ:Inxlg}) the 
concentration scales like $c(t) \sim 1/\sqrt{\pi t}$ for $t \to \infty$. This 
leading order result applies in fact to any DLPC process with $d = 1/2$ and 
arbitrary $\gamma > 0$ and regardless of the initial density $\rho > 0$ \cite{ZhoBen95}. 
To understand why, consider the scaling of the domain size distribution 
Eq.~(\ref{equ:DkI}). In the limit $t \to \infty$ and for domain sizes $k = 
\mathcal{O}(\sqrt{t})$ one finds, via Eqs.~(\ref{equ:Inrec}) and (\ref{equ:Inxnlg}), 
the scaling 
\begin{equation}
  D_k(t) = \frac{k}{t} \, \rme^{-2t} \, I_k(2t) \sim \frac{1}{\sqrt{\pi} \, t} \, 
  \frac{k}{2\sqrt{t}} \, \rme^{-{k^2}/({4t})}. 
  \label{equ:Dkscaling}
\end{equation}
Typical domains have a size of $k=O\left(\sqrt{t}\right)$ and density 
$O\left(t^{-1}\right)$. Integrating Eq.~(\ref{equ:Dkscaling}) over $\kappa 
= k/(2 \sqrt{t}) \in \, ]0,\infty[$ shows that only typical domains 
of length $k=O\left(\sqrt{t}\right)$ contribute to the scaling $c(t) \sim 
1/\sqrt{\pi t}$. Large domains $k \gg \sqrt{t}$ are exponentially unlikely 
and also small domains $k \ll \sqrt{t}$ are of subdominant density. 
In fact, the density of finite domains with $k = \mathcal{O}(1)$ is
$D_k(t) \sim k/(2 \sqrt{\pi} \, t^{3/2}) = \mathcal{O}(t^{-3/2})$,
while for typical domains $D_k(t)=\mathcal{O}(t^{-1})$.
The scaling of 
DLPC is thus controlled by typical domains. For such domains to disappear, the 
delimiting particles first have to diffuse for a time $\mathcal{O}(t)$ 
before there is a significant chance that they will occupy adjacent sites. 
The coagulation reaction itself then occurs on an $\mathcal{O}(1)$ time 
scale for any $\gamma > 0$. It therefore only has a subdominant effect on 
the rate at which typical domains disappear, and the process is said to be in the 
diffusion controlled regime \cite{ZhoBen95}. The scaling for the density of 
typical domains Eq.~(\ref{equ:Dkscaling}) thus applies to any DLPC process 
with $d=1/2$ and $\gamma > 0$, and regardless of the initial density 
$\rho > 0$. The precise value of the coagulation rate $\gamma$ only shows up 
in the coefficient for the density of small domains $k = \mathcal{O}(1)$ 
which, in any case, is $\mathcal{O}(t^{-3/2})$.  

The above discussion illustrates that the initial density $\rho$ as
well as the coagulation rate $\gamma$ only have a subdominant effect
on the evolution of DLPC \cite{ZhoBen95} for $t \gg 1$; recall that in 
the FA model this long-time regime
corresponds to unscaled times much larger than $1/\ceq$ but much
smaller $1/\ceq^2$. The long-time behaviour of the effective FA process
Eq.~(\ref{equ:WFAeff}) can therefore be extracted from the analytically
tractable free fermion case ($d = \gamma = 1/2$). We also exploit the
irrelevance of the initial density for the leading order long-time
results to set $\rho = 1$; this simplifies the mathematics and
produces differences to the case $\rho < 1$ only in subdominant terms.

Note that the above simplifications will hold for all observables 
except for those which show particular sensitivity to the density of small 
domains; we will return to this point in our discussion 
of response functions.

\section{Two-Time Quantities}
\label{sec:2t}

In the following we derive exact scaling expressions for the two-time correlation
and response functions of the FA model in the long-time regime
discussed above, by exploiting the equivalence to the free fermion 
DLPC process with initial density $\rho = 1$.

The mapping from DLPC to DLPA, which is our key for deriving results, cannot be applied 
in a practical way to multi-time quantities. Consider, for instance, 
\begin{equation} 
  \langle A_1(t) A_2(\tw) \rangle_c = \bra{e} A_1 \, \rme^{\Wc \dt} A_2 \rme^{\Wc \tw} \ket{\rho}, 
  \label{equ:mapA1A2} 
\end{equation}
where $\tw$ is the waiting time since preparation of the system
and $\dt=t-\tw$ is the time interval to the later measurement at time $t$.
Inserting identity operators $B B^{-1}$ yields $\bra{e} \tilde{A}_1 \, \rme^{\Wa \dt} \tilde{A}_2 
\rme^{\Wa \tw} \ket{\frac{1}{2}\rho}$. But as discussed below Eq.~(\ref{equ:Aca}) the 
mapped observables $\tilde{A}_{1,2} = B^{-1} A_{1,2} B$ are generally non-diagonal 
operators. While the projection $\bra{e} \tilde{A}_1$ may reduce $\tilde{A}_1$ to 
a physical observable this is no longer the case for $\tilde{A}_2$ which is sandwiched 
between evolution operators. 

We now demonstrate how the mapping can nevertheless be applied to multi-time quantities 
and in a physically sensible manner if the evolution operator is known. The discussion 
will proceed in the reverse order of Sec.~\ref{sec:map}: we first consider spin dynamics 
in the Glauber Ising chain, relate these to parity intervals in DLPA and then map onto 
DLPC. We exploit the following general result for the evolution of two-spin correlations 
$F^{(2)}_{\ivec}(t) = \bra{e} \sigma^z_{i_1} \sigma^z_{i_2} \rme^{\Ws t} \ket{\psi}$ in 
the Glauber Ising model at zero temperature: in the notation of \cite{MaySol04}, 
\begin{equation}
  F^{(2)}_{\ivec}(t) = \sum_{j_1<j_2} G^{(2)}_{\ivec,\jvec}(t) A^{(2)}_{\jvec} + 
  H_{i_2-i_1}(2t) A^{(0)}, 
  \label{equ:spinprop1}
\end{equation}
where
\begin{equation}
  \G{\ivec,\jvec}(t) = \rme^{-2t} \left[ I_{i_1-j_1} I_{i_2-j_2} - 
  I_{i_1-j_2} I_{i_2-j_1} \right](t). 
  \label{equ:G} 
\end{equation}
and the $A^{(2)}_{\jvec} = \bra{e} \sigma^z_{j_1}
\sigma^z_{j_2} \ket{\psi}$ are the correlations at time 0 while
$A^{(0)} = \braket{e}{\psi}$. Eq.~(\ref{equ:spinprop1})  
expresses a two-spin correlation at time $t$ in terms of those in the general 
state $\ket{\psi}$ at $t=0$. The corresponding two-spin  
Green's function $\G{\ivec,\jvec}(t)$ is expressed in 
Eq.~(\ref{equ:G}) in terms of modified Bessel functions $I_n(t)$. The
definition of the latter is recalled in 
Eq.~(\ref{equ:Inint}), while
$H_n(t)$ is stated explicitly in Eq.~(\ref{equ:Hndef}). Equation (\ref{equ:spinprop1}) applies for
$i_1 < i_2$ and it is our convention throughout that sums as in Eq.~(\ref{equ:spinprop1}) 
are taken over ordered integer pairs $\jvec=(j_1, j_2)$.
Now, given that Eq.~(\ref{equ:spinprop1}) applies for arbitrary
initial states $\ket{\psi}$ 
we may equivalently write 
\begin{equation}
  \bra{e} \sigma^z_{i_1} \sigma^z_{i_2} \rme^{\Ws t} = 
  \sum_{j_1<j_2} G^{(2)}_{\ivec,\jvec}(t) \bra{e} \sigma^z_{j_1} \sigma^z_{j_2} + 
  H_{i_2-i_1}(2t) \bra{e}. 
  \label{equ:spinprop}  
\end{equation}
We refer to this object as the two-spin propagator; more general multi-spin propagators 
follow likewise from the general result given in \cite{MaySol04}. Equation (\ref{equ:spinprop}) 
has a direct analogue in DLPA. Recall that according to Eq.~(\ref{equ:Ps}) the operator 
$\sigma_{i_1}^z \sigma_{i_2}^z = P_{i_1,i_2}$ measures the parity of the number of 
domain walls between sites $i_1$ and $i_2$. This, combined with the fact that Glauber 
dynamics of the spin system correspond to DLPA of the domain walls, implies 
\begin{equation}
  \bra{e} P_{\ivec} \, \rme^{\Wa t} = 
  \sum_{j_1<j_2} G^{(2)}_{\ivec,\jvec}(t) \bra{e} P_{\jvec} + 
  H_{i_2-i_1}(2t) \bra{e}, 
  \label{equ:parityprop}  
\end{equation}
which is the parity propagator in DLPA. A mapping to DLPC is now straightforward. 
Multiplying by $B^{-1}$ from the right and inserting identity operators $B^{-1} B$ 
immediately yields the empty interval propagator in DLPC, 
\begin{equation}
  \bra{e} E_{\ivec} \, \rme^{\Wc t} = 
  \sum_{j_1<j_2} G^{(2)}_{\ivec,\jvec}(t) \bra{e} E_{\jvec} + 
  H_{i_2-i_1}(2t) \bra{e}.
  \label{equ:emptyprop}
\end{equation}
The expressions Eqs.~(\ref{equ:spinprop}--\ref{equ:emptyprop}) are rather remarkable. 
They show that propagating a two-spin operator forward in time in the Glauber Ising chain 
is equivalent to propagating a parity operator in DLPA or an empty interval operator 
in DLPC. Nevertheless, as we will see, the two-time quantities that follow from 
Eqs.~(\ref{equ:parityprop}) and (\ref{equ:emptyprop}) are not related in a simple way. 
Note also that in contrast to Eq.~(\ref{equ:mapA1A2}) no non-diagonal
operators arise explicitly when the mapping is written in terms of the
propagators: since Eqs.~(\ref{equ:parityprop}) and (\ref{equ:emptyprop}) 
are explicit representations of the evolution operator, a projection state $\bra{e}$ is 
always available to absorb contributions from off-diagonal components
of mapped operators. We finally add that 
via the identity Eq.~(\ref{equ:Hnsum}) each of the above propagators can be reexpressed 
in a somewhat different but equivalent form, for instance 
\begin{equation}
  \bra{e} E_{\ivec} \, \rme^{\Wc t} = \bra{e} + \sum_{j_1 < j_2} \G{\ivec,\jvec}(t) \, 
  \bra{e} \left( E_{\jvec} - 1 \right). 
  \label{equ:emptypropnoH} 
\end{equation}

\subsection{Two-Time Correlations}
\label{sec:cor2t}

Using the DLPC propagator derived above we can now analyse the connected two-time 
correlation functions of mobility excitations (or defects) in the FA model,
\begin{equation}
  C_{i-j}(t,\tw) = \bra{e} \n_i \, \rme^{\Wc \dt} \, \n_j \, \rme^{\Wc \tw} \ket{1} - 
  \bra{e} \n_i \, \rme^{\Wc t} \ket{1} \bra{e} \n_j \, \rme^{\Wc \tw} \ket{1}.
  \label{equ:cor2tdef}
\end{equation}
To understand nonequilibrium fluctuation-dissipation relations one also
requires the associated response functions; these will be considered
in the next subsection. It is useful to introduce the more general correlator 
\begin{equation} 
  C_{\ivec,\jvec}(t,\tw) = \bra{e} E_{\ivec} \, \rme^{\Wc \dt} \, E_{\jvec} \, 
  \rme^{\Wc \tw} \ket{1} - \bra{e} E_{\ivec} \, \rme^{\Wc t} \ket{1} 
  \bra{e} E_{\jvec} \, \rme^{\Wc \tw} \ket{1}, 
  \label{equ:cor2tE}
\end{equation}
which reduces to $C_{i-j}(t,\tw)$ for $\ivec=(i,i+1)$ and $\jvec=(j,j+1)$ 
since $\n_i = 1 - E_{i,i+1}$. The first step in evaluating Eq.~(\ref{equ:cor2tE}) consists 
in decomposing the evolution operator $\rme^{\Wc t} = \rme^{\Wc \dt} \, \rme^{\Wc \tw}$ 
and substituting the empty interval propagator
Eq.~(\ref{equ:emptypropnoH}) for both occurrences of 
$\bra{e} E_{\ivec} \, \rme^{\Wc \dt}$ in Eq.~(\ref{equ:cor2tE}). Simplifications occur 
due to probability conservation, $\bra{e} \rme^{\Wc \dt} = \bra{e}$,
and the normalization $\braket{e}{1} = 1$ of the initial state, giving 
\begin{equation} 
  \fl C_{\ivec,\jvec}(t,\tw) = \sum_{k_1 < k_2} \G{\ivec,\kvec}(\dt) \left( \bra{e} 
  E_{\kvec} E_{\jvec} \, \rme^{\Wc \tw} \ket{1} - \bra{e} E_{\kvec} \, \rme^{\Wc \tw} \ket{1} 
  \bra{e} E_{\jvec} \, \rme^{\Wc \tw} \ket{1} \right). 
  \label{equ:cor2tE1} 
\end{equation} 
The connected two-time empty interval correlations are now reduced to connected one-time averages. 
We remark that precisely the same expression applies in DLPA if one replaces the empty 
by parity interval operators. Nevertheless, as indicated above, the resulting two-time 
correlations are different. This is for the following reason: the sum in Eq.~(\ref{equ:cor2tE1}) 
runs over all ordered index pairs $\kvec$ and thus there are inevitably overlaps of $E_{\kvec}$ 
with $E_{\jvec}$. In such cases the empty interval operators merge since $(1-\n_i)^2 = 1-\n_i$. 
Parity interval operators, on the other hand, cancel each other in regions of overlap because 
$(1-2\n_i)^2 = 1$. Differences in two-time correlations between DLPC and DLPA therefore arise 
due to the different reduction properties of empty and parity interval operators. For the case 
of empty intervals that we are interested in here one has 
\begin{equation}
  E_{k_1,k_2} E_{j_1,j_2} = 
  \left\{
  \begin{array}{l}
    E_{k_1,k_2} E_{j_1,j_2} \\
    E_{k_1,j_2} \\
    E_{k_1,k_2} \\
    E_{j_1,j_2} \\
    E_{j_1,k_2} \\
    E_{j_1,j_2} E_{k_1,k_2} 
  \end{array}
  \quad \mbox{for} \quad 
  \begin{array}{c} 
    k_1 < k_2 < j_1 < j_2 \\ 
    k_1 < j_1 \leq k_2 \leq j_2 \\ 
    k_1 < j_1 < j_2 < k_2 \\ 
    j_1 \leq k_1 < k_2 \leq j_2 \\ 
    j_1 \leq k_1 \leq j_2 < k_2 \\ 
    j_1 < j_2 < k_1 < k_2 
  \end{array}
  \right..
  \label{equ:hierarchy}
\end{equation}
To proceed with the derivation of $C_{\ivec,\jvec}(t,\tw)$ the summation in Eq.~(\ref{equ:cor2tE1}) over 
the pair-correlation term must be broken up according to Eq.~(\ref{equ:hierarchy}), with corresponding 
reductions in each sector. For the remaining averages, which are now of the form $\bra{e} E_{l_1,l_2} \, 
\rme^{\Wc \tw} \ket{1}$ and $\bra{e} E_{l_1,l_2} E_{l_3,l_4} \, \rme^{\Wc \tw} \ket{1}$ with strictly 
ordered indices $l_1 < l_2 < l_3 < l_4$, we substitute the results Eqs.~(\ref{equ:E1t}) and 
(\ref{equ:E2t}) from above. At this point one has a rather bulky expression for $C_{\ivec,\jvec}(t,\tw)$. 
To simplify matters we focus on $\jvec=(j,j+1)$ -- the case of interest -- which narrows down several 
summation ranges, c.f.\ Eq.~(\ref{equ:hierarchy}). It is then a tedious but trivial task to show that 
the various sums can be regrouped into the compact form 
\begin{eqnarray}
\fl C_{\ivec,(j,j+1)}(t,\tw) = \left[ 1- H_1(2\tw) \right] \sum_{k_1 \leq j} \sum_{k_2 > j} 
    \G{\ivec,\kvec}(\dt) H_{k_2-k_1}(2\tw) 
  \nonumber \\ 
  + \sum_{k_1 \leq j} \sum_{k_2 \leq j} 
    \G{\ivec,\kvec}(\dt) H_{j+1-k_1}(2\tw) \left[ \delta_{k_2,j} + H_{j-k_2}(2\tw) \right] 
  \nonumber \\ 
  + \sum_{k_1 > j} \sum_{k_2 > j} 
    \G{\ivec,\kvec}(\dt) H_{k_2-j}(2\tw) \left[ \delta_{k_1,j+1} + H_{k_1-j-1}(2\tw) \right]. 
  \label{equ:cor2tE2} 
\end{eqnarray}
Here antisymmetry of $\G{\ivec,\kvec}(\dt)$ under permutation of $k_1 \leftrightarrow k_2$ and 
$H_0(2\tw) = 0$ was used. Finally we set $\ivec = (i,i+1)$ and rearrange Eq.~(\ref{equ:cor2tE2}) 
as follows: in each line the summation variables are changed from $k_1,k_2$ to $p,q \geq 0$ 
and the indices of the Green's functions are simplified based on translational $\G{\ivec,\kvec}(\dt) = \G{\ivec+\lvec,\kvec+\lvec}(\dt)$ and reflection $\G{\ivec,\kvec}(\dt) = \G{-\ivec,-\kvec}(\dt)$ 
symmetry. In terms of $n=i-j$ the last two lines of Eq.~(\ref{equ:cor2tE2}) thereby become 
\begin{equation}
  \sum_{p,q=0}^\infty \G{(\pm n,\pm n-1),(p,q)}(\dt) \, H_{p+1}(2\tw) \left[ \delta_{q,0} 
  + H_q(2\tw) \right], 
  \label{equ:cancel} 
\end{equation}
where the two signs $+n$ and $-n$ correspond to the third and second line of 
Eq.~(\ref{equ:cor2tE2}), respectively. 
For later convenience we point out that hidden in this expression lies a leading order cancellation 
in the limit of large $\tw$. Using that the Green's function in
Eq.~(\ref{equ:cancel}) changes sign when $p$ and $q$ are interchanged
we can write
\begin{equation}
  \sum_{p,q=0}^\infty \G{(\pm n,\pm n-1),(p,q)}(\dt) \, \left[ \delta_{p,0} + H_p(2\tw) \right] 
  \left[ \delta_{q,0} + H_q(2\tw) \right] = 0.
  \label{equ:cancel1} 
\end{equation}
Subtracting this vanishing expression from Eq.~(\ref{equ:cancel}) replaces $H_{p+1}(2\tw)$ 
with $H_{p+1}(2\tw)-\delta_{p,0}-H_p(2\tw) = -\rme^{-2\tw} [I_p + I_{p+1}](2\tw)$, see 
Eq.~(\ref{equ:Hnrec}). Similarly, in the first line of Eq.~(\ref{equ:cor2tE2}) we have 
$1-H_1(2\tw) = \rme^{-2\tw} [I_0 + I_1](2\tw)$. Altogether our result for the connected 
two-time mobility correlation function defined in Eq.~(\ref{equ:cor2tdef}) becomes 
\begin{eqnarray}
\fl C_n(t,\tw) = 
	\rme^{-2\tw} [I_0 + I_1](2\tw) \sum_{p,q=0}^\infty \G{(n,n+1),(-p,q+1)}(\dt) \, H_{p+q+1}(2\tw) 
  \nonumber \\ 
  - \sum_{p,q=0}^\infty \G{(+n,+n-1),(p,q)}(\dt) \, \rme^{-2\tw} [I_p + I_{p+1}](2\tw) 
  \left[ \delta_{q,0} + H_q(2\tw) \right] 
  \nonumber \\ 
  - \sum_{p,q=0}^\infty \G{(-n,-n-1),(p,q)}(\dt) \, \rme^{-2\tw} [I_p + I_{p+1}](2\tw) 
  \left[ \delta_{q,0} + H_q(2\tw) \right]. 
\label{equ:cor2tsums}
\end{eqnarray}
This expression is exact for free-fermion DLPC with unit initial
density, but as argued above its scaling behaviour at large $\tw$ applies to
DLPC processes with arbitrary coagulation rate and initial density. This includes in particular the effective 
FA dynamics we are interested in. Equation~(\ref{equ:cor2tsums}) forms the basis for our 
subsequent discussion; the leading order cancellation with respect 
to $\tw$ having been eliminated, it is well suited for long-time expansions.

\subsection{Two-Time Response Functions} 
\label{sec:res2t} 

We now turn to the two-time response function conjugate to the two-time correlation $C_n(t,\tw)$ 
discussed above, which is given by 
\begin{equation}
  R_{i-j}(t,\tw) = \bra{e} \n_i \, \rme^{\Wc \dt} \, V_j \, \rme^{\Wc \tw} \ket{1}, 
  \label{equ:res2tdef}
\end{equation}
where the operator $V_j = T \left[ \partial \Wc(h)/\partial h \right]|_{h=0}$ accounts 
for the perturbation $\delta \mathcal{H} = -h \, \n_j$ of the effective dynamics at $\tw$. 
Note that the response contains a factor of $T$ consistent with our definition Eq.~(\ref{equ:R2tdef}). 
By analogy to the correlations it is convenient to consider the more general response 
\begin{equation}
  R_{\ivec,j}(t,\tw) = - \bra{e} E_{\ivec} \, \rme^{\Wc \dt} \, 
  V_j \, \rme^{\Wc \tw} \ket{1}, 
  \label{equ:res2tE}
\end{equation}
which reduces to Eq.~(\ref{equ:res2tdef}) for $\ivec=(i,i+1)$ as is obvious from $E_{i,i+1} = 1 - \n_i$ 
and probability conservation $\bra{e} \rme^{\Wc \dt} V_j = \bra{e} V_j = 0$. 
As before, substitution of the empty interval propagator Eq.~(\ref{equ:emptypropnoH}) reduces the 
two-time average in Eq.~(\ref{equ:res2tE}) to a one-time average, which is most conveniently 
expressed in the form 
\begin{equation}
  R_{\ivec,j}(t,\tw) = \sum_{k_1 < k_2} \G{\ivec,\kvec}(\dt) R_{\kvec,j}(\tw,\tw), 
  \label{equ:res2tG} 
\end{equation} 
with the instantaneous response 
\begin{equation} 
  R_{\ivec,j}(\tw,\tw) = - \bra{e} E_{\ivec} V_j \, \rme^{\Wc \tw} \ket{1}. 
  \label{equ:res1t}
\end{equation}
In what follows we will first concentrate on the instantaneous response; 
from this the two-time response $R_{\ivec,j}(t,\tw)$ then follows immediately via 
Eq.~(\ref{equ:res2tG}). The effect of the perturbation encoded in $V_j$ 
has the explicit form 
\begin{equation}
  V_j = \frac{1}{2} (F_{j-1} F_j - 1) \n_{j-1} (1-\n_j) + 
  \frac{1}{2} (F_{j+1} F_j - 1) \n_{j+1} (1-\n_j).
  \label{equ:V}
\end{equation}
Here the first term is nonzero only if site $j$ is empty and site
$j-1$ occupied, reflecting the increased rate for diffusion from
$j-1$ to $j$ as discussed at the end of Sec.~\ref{sec:effective}; the second term similarly captures 
the increased diffusion from $j+1$ to $j$. It will become clear below that a 
decomposition of $V_j$ into an asymmetric and symmetric contribution, 
$V_j = \Va_j + \Vs_j$, is very useful: 
\begin{eqnarray}
  \Va_j = \frac{1}{4} (F_{j-1} F_j - 1) 
  (\n_{j-1} - \n_j) \hspace{1ex} + 
   \frac{1}{4} (F_{j+1} F_j - 1) 
  (\n_{j+1} - \n_j), 
  \label{equ:Va} \\ 
  \Vs_j = \frac{1}{4} (F_{j-1} F_j - 1) 
  (\n_{j-1} - \n_j)^2 + 
   \frac{1}{4} (F_{j+1} F_j - 1) 
  (\n_{j+1} - \n_j)^2. 
  \label{equ:Vs} 
\end{eqnarray}
The operator $\Va_j$ represents increased local diffusion rates $j-1 \rightarrow j 
\leftarrow j+1$ for entering site $j$ but decreased ones for leaving it 
$j-1 \leftarrow j \rightarrow j+1$, thus $\Va_j$ {\em traps diffusing defects} at site $j$. 
Its symmetric counter-part $\Vs_j$, on the other hand, corresponds to increased local 
diffusion rates between sites $j-1 \leftrightharpoons j \leftrightharpoons j+1$ in both 
directions and hence {\em speeds up the local dynamics}. Accordingly, the defect response 
function in the FA model may be decomposed into the contributions 
\begin{equation}
  R_n(t,\tw) = \ra_n(t,\tw) + \rs_n(t,\tw). 
  \label{equ:res2tas}
\end{equation}

\subsubsection{Asymmetric Perturbation}
\label{sec:res2ta} 

We first discuss the effect of the asymmetric part of the
perturbation. From Eq.~(\ref{equ:res1t})
the corresponding instantaneous response function is 
\begin{equation}
  \ra_{\ivec,j}(\tw,\tw) = -\bra{e} E_{\ivec} \Va_j \rme^{\Wc \tw} \ket{1}. 
  \label{equ:res1tadef}
\end{equation}
If $\Va_j$ acts entirely inside or outside of 
$E_{\ivec}$, then the instantaneous response vanishes: the empty interval density 
$\bra{e} E_{\ivec} \, \rme^{\Wc \tw} \ket{1}$ is only affected by the perturbation 
when applied at either boundary of $E_{\ivec}$, increasing/decreasing the rate 
for defects to enter/exit the interval. Substituting the expressions (\ref{equ:EPdef}) 
and (\ref{equ:Va}) for $E_{\ivec}$ and $\Va_j$, respectively, and using the 
commutation relation $(1 - \n_i) F_i = F_i \n_i$ one 
shows that 
\begin{eqnarray}
\fl \ra_{\ivec,j}(\tw,\tw) = \frac{1}{4} \left( \delta_{j,i_1} - \delta_{j,i_1 - 1} \right) 
  \bra{e} \left( \n_{i_1 - 1} - \n_{i_1} \right)^2 E_{i_1 + 1, i_2} \, \rme^{\Wc \tw} \ket{1}
  \nonumber \\ 
  - \frac{1}{4} \left( \delta_{j,i_2} - \delta_{j,i_2 - 1} \right) 
  \bra{e} E_{i_1, i_2 - 1} \left( \n_{i_2 - 1} - \n_{i_2} \right)^2  \rme^{\Wc \tw} \ket{1}. 
  \label{equ:res1taE} 
\end{eqnarray}
This makes perfect sense: the Kronecker deltas enforce that $j$ must be a relevant 
site and also carry the sign of the response, while the averages give its magnitude. 
In the first line, for instance, only the configurations $0_{i_1-1} 1_{i_1} 0_{i_1+1} 
\ldots 0_{i_2-1}$ and $1_{i_1-1} 0_{i_1} 0_{i_1+1} \ldots 0_{i_2-1}$ contribute, since 
these are susceptible to perturbations of the local diffusion rates between sites $i_1-1$ 
and $i_1$ and affect the empty interval observable $E_{\ivec}$. To evaluate the 
averages it is convenient to rewrite $\left( \n_{i_1 - 1} - \n_{i_1} \right)^2 = 
\n_{i_1 - 1} \left( 1 - \n_{i_1} \right) + \n_{i_1}  - \n_{i_1 - 1} \n_{i_1}$ whereby 
\begin{equation}
  \left( \n_{i_1 - 1} - \n_{i_1} \right)^2 E_{i_1+1, i_2} = \n_{i_1-1} E_{i_1,i_2} + 
  \n_{i_1} E_{i_1+1,i_2} - \n_{i_1 - 1} \n_{i_1} E_{i_1+1,i_2}. 
  \label{equ:Eaid} 
\end{equation}
The first two terms on the right hand side of this equation are related to the density 
of domains: the observable $\n_{i_1} E_{i_1+1,i_2}$, for instance, measures the density 
of domains $1_{i_1} 0_{i_1+1} \ldots 0_{i_1+n-1} 1_{i_1+n}$ of any size $n \geq i_2 - i_1$ 
and hence 
\begin{equation}
  \bra{e} \n_{i_1} E_{i_1+1,i_2} \, \rme^{\Wc \tw} \ket{1} = \sum_{k=0}^\infty D_{i_2-i_1+k}(\tw). 
  \label{equ:nED} 
\end{equation}
Altogether we may rewrite Eq.~(\ref{equ:res1taE}) in the form 
\begin{equation}
  \ra_{\ivec,j}(\tw,\tw) = \frac{1}{4} A_{\ivec,j}  
  \left[ D_{i_2-i_1}(\tw) + 2 \sum_{k=1}^\infty D_{i_2-i_1+k}(\tw) \right] 
  + \Da_{\ivec,j}(\tw,\tw), 
  \label{equ:res1taD} 
\end{equation}
where $A_{\ivec,j} = \delta_{j,i_1} - \delta_{j,i_1-1} - \delta_{j,i_2} + \delta_{j,i_2-1}$ and 
\begin{eqnarray}
\fl \Da_{\ivec,j}(\tw,\tw) = -\frac{1}{4} \left( \delta_{j,i_1} - \delta_{j,i_1-1} \right) 
  \bra{e} \n_{i_1-1} \n_{i_1} E_{i_1+1,i_2} \, \rme^{\Wc \tw} \ket{1}
  \nonumber \\ 
  +\frac{1}{4} \left( \delta_{j,i_2} - \delta_{j,i_2-1} \right) 
  \bra{e} E_{i_1,i_2-1} \n_{i_2-1} \n_{i_2} \, \rme^{\Wc \tw} \ket{1}. 
  \label{equ:Deltaa} 
\end{eqnarray}
The function $\Da_{\ivec,j}(\tw,\tw)$ in Eq.~(\ref{equ:res1taD}) was isolated since it 
is subdominant at large $\tw$ and thus unimportant for the long-time scaling. The
remaining terms on the r.h.s.\ of
Eq.~(\ref{equ:res1taD}) contain a sum over the density of domains $D_n(\tw)$ which picks 
up its main contributions from domains of size $n =
\mathcal{O}(\sqrt{\tw})$.
In this regime $D_n(\tw)$ has a robust scaling form
Eq.~(\ref{equ:Dkscaling}) that is independent of the coagulation rate
and the same therefore applies to the leading long-time behaviour of
the instantaneous asymmetric response.
It is clear 
from Eq.~(\ref{equ:Eaid}) that $\Da_{\ivec,j}(\tw,\tw)$ corrects for the fact that, e.g.,  
the configuration $1_{i_1-1} 1_{i_1} 0_{i_1+1} \ldots 0_{i_2-1}$ does not respond to perturbations 
of the local diffusion rates between sites $i_1-1$ and $i_1$. At long
times such
states with neighbouring defects have subdominant density in DLPC
compared to states with isolated defects, and this is the intuitive 
reason why $\Da_{\ivec,j}(\tw,\tw)$, whose scaling does depend on the
precise coagulation rate, is subdominant compared to the leading term
in Eq.~(\ref{equ:res1taD}). (In the effective FA dynamics the
probability of states containing adjacent defects is further suppressed
by the fact that coagulation is essentially instantaneous.)
Further discussion of $\Da_{\ivec,j}(\tw,\tw)$ 
may be found in \ref{sec:delta}; here we focus on Eq.~(\ref{equ:res1taD}) which controls 
the asymptotic long-time scaling. Substituting the DLPC result
Eq.~(\ref{equ:DkI}) for $D_n(\tw)$ produces a telescopic sum
in Eq.~(\ref{equ:res1taD}) and gives
\begin{equation}
\fl \ra_{\ivec,j}(\tw,\tw) = \frac{1}{4} A_{\ivec,j} \, 
  \rme^{-2\tw} [I_{i_2-i_1-1} + 2 I_{i_2-i_1} + I_{i_2-i_1+1}](2\tw) 
  + \Da_{\ivec,j}(\tw,\tw). 
  \label{equ:res1taI} 
\end{equation}
The associated two-time response function is readily obtained via Eq.~(\ref{equ:res2tG}). 
The ordered double sum over $k_1 < k_2$ reduces to semi-infinite ones due to $A_{\ivec,j}$ 
but the latter may be recombined into full summations over all integers and evaluated using 
the convolution property Eq.~(\ref{equ:Inconv}). The resulting expression for the asymmetric 
part of the two-time response function Eq.~(\ref{equ:res2tdef}) with $n=i-j$ is most 
compactly expressed as 
\begin{eqnarray}
\fl  \ra_n(t,\tw) = \partial_{\tw} \frac{1}{2} \rme^{-2t} I_n(t-\tw) [ I_{n-1} 
  + 2 I_n + I_{n+1} ](t+\tw) + \Da_n(t,\tw). 
  \label{equ:res2ta} 
\end{eqnarray}

\subsubsection{Symmetric Perturbation}
\label{sec:res2ts} 

The derivation of the response to a symmetric perturbation is largely analogous to 
the asymmetric case discussed above. We define
\begin{equation}
  \rs_{\ivec,j}(\tw,\tw) = -\bra{e} E_{\ivec} \Vs_j \, \rme^{\Wc \tw} \ket{1}. 
  \label{equ:res2tsdef}
\end{equation}
Again a non-zero response occurs only if the perturbation is applied at either 
boundary of the empty interval $E_{\ivec}$. Specifically one finds 
\begin{eqnarray}
\fl \rs_{\ivec,j}(\tw,\tw) = 
  \frac{1}{4} \left( \delta_{j,i_1} + \delta_{j,i_1 - 1} \right) 
  \bra{e} \left( \n_{i_1 - 1} - \n_{i_1} \right) E_{i_1 + 1, i_2} \, \rme^{\Wc \tw} \ket{1} 
  \nonumber \\ 
  + \frac{1}{4} \left( \delta_{j,i_2} + \delta_{j,i_2 - 1} \right) 
  \bra{e} E_{i_1, i_2 - 1} \left( \n_{i_2 - 1} - \n_{i_2} \right)  \, \rme^{\Wc \tw} \ket{1}, 
  \label{equ:res1tsE} 
\end{eqnarray}
which is the symmetric counterpart to Eq.~(\ref{equ:res1taE}). This time we rewrite 
$\n_{i_1 - 1} - \n_{i_1} = \n_{i_1 - 1} \left( 1 - \n_{i_1} \right) - \n_{i_1}  + 
\n_{i_1 - 1} \n_{i_1}$ so that 
\begin{equation}
  \left( \n_{i_1 - 1} - \n_{i_1} \right) E_{i_1+1, i_2} = \n_{i_1-1} E_{i_1,i_2} - 
  \n_{i_1} E_{i_1+1,i_2} + \n_{i_1 - 1} \n_{i_1} E_{i_1+1,i_2}, 
  \label{equ:Esid}
\end{equation}
and once more use Eq.~(\ref{equ:nED}) to express the averages $\bra{e} \n_{i_1} E_{i_1+1,i_2} \, 
\rme^{\Wc \tw} \ket{1}$ etc.\ in terms of domain densities $D_n(\tw)$. One finds 
\begin{equation}
  \rs_{\ivec,j}(\tw,\tw) = - \frac{1}{4} B_{\ivec,j}  
  D_{i_2-i_1}(\tw) + \Ds_{\ivec,j}(\tw,\tw), 
  \label{equ:res1tsD}
\end{equation}
where $B_{\ivec,j} = \delta_{j,i_1} + \delta_{j,i_1-1} + \delta_{j,i_2} + \delta_{j,i_2-1}$ and 
\begin{eqnarray}
\fl \Ds_{\ivec,j}(\tw,\tw) = + \frac{1}{4} \left( \delta_{j,i_1} + \delta_{j,i_1-1} \right) 
  \bra{e} \n_{i_1-1} \n_{i_1} E_{i_1+1,i_2} \, \rme^{\Wc \tw} \ket{1}
  \nonumber \\ 
  +\frac{1}{4} \left( \delta_{j,i_2} + \delta_{j,i_2-1} \right) 
  \bra{e} E_{i_1,i_2-1} \n_{i_2-1} \n_{i_2} \, \rme^{\Wc \tw} \ket{1}. 
  \label{equ:Deltas}
\end{eqnarray}
Due to the difference in signs between Eqs.~(\ref{equ:Eaid}) and (\ref{equ:Esid}) we end up 
with a single domain density $D_{i_2-i_1}(\tw)$ in the symmetric response Eq.~(\ref{equ:res1tsD}) 
rather than the sum we had in the asymmetric case Eq.~(\ref{equ:res1taD}). This has important 
consequences for the scaling behaviour of $\rs_{\ivec,j}(\tw,\tw)$. The excess term 
$\Ds_{\ivec,j}(\tw,\tw)$, Eq.~(\ref{equ:Deltas}), differs from $\Da_{\ivec,j}(\tw,\tw)$, 
Eq.~(\ref{equ:Deltaa}), only in overall signs; we therefore discuss both
quantities together in \ref{sec:delta}.
At fixed $i_2-i_1$ and large $\tw$ we have, for instance, that both 
$\Da_{\ivec,j}(\tw,\tw)$ and $\Ds_{\ivec,j}(\tw,\tw)$ scale like $\mathcal{O}\left(1/{\tw}^{3/2}\right)$. 
This is a subdominant contribution to $\ra_{\ivec,j}(\tw,\tw) =
\mathcal{O}\left(1/{\tw}^{1/2}\right)$, while 
in $\rs_{\ivec,j}(\tw,\tw) = \mathcal{O}\left(1/{\tw}^{3/2}\right)$ from Eq.~(\ref{equ:res1tsD}) 
it contributes to leading order. Therefore the scaling of $\rs_{\ivec,j}(\tw,\tw)$ for small 
$i_2 - i_1 \geq 1$ and large $\tw$ depends on the precise coagulation
rate and so is different in the effective FA model and e.g.\ free
fermion DLPC.
Only in the regime $i_2 - i_1 \gg 1$ does $\rs_{\ivec,j}(\tw,\tw)$ assume a robust scaling form that 
is independent of the underlying coagulation rate. This is because the density of domains 
$D_n(\tw)$ has a robust scaling in this regime while at the same time $\Ds_{\ivec,j}(\tw,\tw)$ 
becomes negligible in comparison (see \ref{sec:delta}). Having clarified the range of validity of Eq.~(\ref{equ:res1tsD}) 
we now substitute the free fermion DLPC result Eq.~(\ref{equ:DkI}) for the domain density, 
\begin{equation}
  \rs_{\ivec,j}(\tw,\tw) = - \frac{1}{4} B_{\ivec,j} 
  \rme^{-2\tw} [I_{i_2-i_1-1} - I_{i_2-i_1+1}](2\tw) 
  + \Ds_{\ivec,j}(\tw,\tw). 
  \label{equ:res1tsI}
\end{equation}
Inserting this expression into Eq.~(\ref{equ:res2tG}) produces the associated two-time 
response function. As for the asymmetric case the $k$-summations may be carried out using 
the convolution property Eq.~(\ref{equ:Inconv}) of the modified Bessel functions. The 
symmetric part of the two-time response function Eq.~(\ref{equ:res2tdef}) with $n=i-j$ 
becomes 
\begin{eqnarray}
\fl \rs_n(t,\tw) = - \frac{1}{4} \rme^{-2t} \left\{ I_n(t-\tw) [ -I_{n-2} + 2 I_n - I_{n+2} ](t+\tw) \right. 
  \nonumber \\
  \left. + [I_{n-1} - I_{n+1}](t-\tw) [I_{n-1} - I_{n+1}](t+\tw) \right\} + \Ds_n(t,\tw). 
  \label{equ:res2ts} 
\end{eqnarray}
This result is exact for free fermion DLPC; its leading order scaling
at large $\tw$ applies more generally for arbitrary coagulation rates
as long as $\dt \gg 1$. The latter holds true since for
$\dt \gg 1$ the Green's function in Eq.~(\ref{equ:res2tG}) picks up
its leading contributions from $k_2-k_1 = \mathcal{O}(\sqrt{\dt})$, and in this regime
Eq.~(\ref{equ:res1tsI}) has a robust long-time scaling form.

\section{Local Correlation and Response}
\label{sec:local_fdt}

In this section we study the nonequilibrium FDT in the FA model associated with 
the local defect observable $A = \n_i$. Before presenting results for
the FDR we discuss the dynamics of the corresponding connected
two-time local defect autocorrelation  
and response functions, denoted by $\cord(t,\tw)$ and $\resd(t,\tw)$, respectively.

\subsection{Defect Autocorrelation}
\label{sec:cord}

The defect autocorrelation function $\cord(t,\tw) = \langle \n_i(t) \n_i(\tw)\rangle 
-\langle \n_i(t) \rangle \langle \n_i(\tw)\rangle = C_0(t,\tw)$ is given by 
Eq.~(\ref{equ:cor2tsums}) evaluated for $n=0$, 
\begin{eqnarray}
\fl \cord(t,\tw) = 
	\rme^{-2\tw} [I_0 + I_1](2\tw) \sum_{p,q=0}^\infty \G{(0,1),(-p,q+1)}(\dt) \, H_{p+q+1}(2\tw) 
  \nonumber \\ 
  - 2 \sum_{p,q=0}^\infty \G{(0,-1),(p,q)}(\dt) \, \rme^{-2\tw} [I_p + I_{p+1}](2\tw) 
  \left[ \delta_{q,0} + H_q(2\tw) \right]. 
  \label{equ:cord} 
\end{eqnarray}
Equation (\ref{equ:cord}) is an exact expression for the connected
two-time particle autocorrelation in the free fermion DLPC process
with filled initial state. Asymptotically, for
$\tw\to\infty$, it applies for DLPC processes with
arbitrary coagulation rate and homogeneous initial states with finite
particle density. We add that Eq.~(\ref{equ:cord}) is then exact to
leading order in $\tw$ for arbitrary $\dt \geq 0$, i.e.\ including
small $\dt=O(1)$. This is because at large $\tw$, the correlation
function for $\dt$ of $O(1)$ is to leading order unaffected by
coagulation processes, the typical distances of $O(\sqrt{\tw})$ between
particles being too large for them to meet. For $\dt=O(\tw)$, on the other hand,
coagulation events do play a role but the precise extent of their
$O(1)$ durations is irrelevant. Either way, the dynamics of the DLPC 
process is essentially controlled by diffusion and independent of the
coagulation rate \cite{ZhoBen95}.

Let us now discuss some limiting cases of Eq.~(\ref{equ:cord}). For
$\dt=0$, one has $\G{\ivec,\jvec}(0) = \delta_{i_1-j_1} 
\delta_{i_2-j_2} - \delta_{i_1-j_2} \delta_{i_2-j_1}$ and so only the $p=q=0$ term 
from the first sum remains while the second one vanishes altogether. Using 
$H_1(2\tw) = 1 - \rme^{-2\tw} [I_0 + I_1](2\tw)$, see Eq.~(\ref{equ:Hnrec}), 
we obtain
\begin{equation}
  \cord(\tw,\tw) = c(\tw) - c^2(\tw), 
  \label{equ:cordcc} 
\end{equation}
with $c(\tw) = \langle \n_i(\tw) \rangle$ the defect concentration given in 
Eq.~(\ref{equ:cn}). This result is consistent with the definition of
$\cord(t,\tw)$ as it must be.
In the quasi-equilibrium regime $\dt \geq 0$ fixed and 
$\tw \to \infty$, convergence of the sums in Eq.~(\ref{equ:cord}) 
is guaranteed by the Green's functions, with relevant contributions originating 
from $p,q = \mathcal{O}(1)$. We may therefore substitute the asymptotic 
expansions Eqs.~(\ref{equ:Inxlg}) and (\ref{equ:Hnxlg}) for the $\tw$-dependent 
functions, 
\begin{equation}
  \cord(t,\tw) \sim \frac{1}{\sqrt{\pi \tw}} \sum_{p,q=0}^\infty \left[ 
  \G{(0,1),(-p,q+1)}(\dt) - 2 \G{(0,-1),(p,q)}(\dt) \right]. 
  \label{equ:cordstG} 
\end{equation}
After substitution of Eq.~(\ref{equ:G}) for the Green's functions the double 
sum factorizes. There are several cancellations and only one sum of the type 
Eq.~(\ref{equ:Ingen}) remains. We find 
\begin{equation}
  \cord(t,\tw) \sim \frac{1}{\sqrt{\pi \tw}} \rme^{-\dt} I_0(\dt) \sim c(\tw) p_r(\dt). 
  \label{equ:cordst}
\end{equation}
Here we have identified the probability of return to the origin for a
standard random walker, $p_r(\tau) = \rme^{-\tau} \, I_0(\tau)$. 
The leading contribution to the two-time correlation arises from 
$\cord(t,\tw) \sim \langle \n_i(t) \n_i(\tw) \rangle$ and its value 
is thus given by the probability $c(\tw)$ of having a defect at site 
$i$ at time $\tw$ multiplied by the probability $p_r(t-\tw)$ for 
this defect to return and occupy the same site at time $t$. 
This reasoning confirms the intuition that in the
quasi-equilibrium regime defects
are too far from each other to meet and react and so behave as isolated random
walkers; this argument of course applies to both DLPC and DLPA \cite{MayBerGarSol03,MaySol05}. 

Finally we consider the aging regime 
$\dt,\tw \gg 1$ with the time ratio $\dt/\tw$ fixed and finite. 
Here asymptotic expansions of the Green's functions
in Eq.~(\ref{equ:cord}) are required, 
\begin{eqnarray}
  \G{(0,1),(-p,q+1)}(\dt) & = & \rme^{-2\dt} \left[ I_p I_q - I_{p+1} I_{q+1} \right](\dt), 
  \label{equ:Gcord1} \\
  \G{(0,-1),(p,q)}(\dt)   & = & \rme^{-2\dt} \left[ I_p I_{q+1} - I_{p+1} I_q \right](\dt). 
  \label{equ:Ccord2} 
\end{eqnarray}
Some care has to be taken since the leading contributions of the modified Bessel 
functions cancel. To obtain the first subdominant correction we rewrite 
\begin{equation} 
  \fl I_{p+1}(\dt) = \frac{1}{2} [ I_{p-1} + I_{p+1} ](\dt) - \frac{1}{2} [ I_{p-1} - I_{p+1} ](\dt) 
  = \partial_{\dt} I_p(\dt) - \frac{p}{\dt} I_p(\dt), 
  \label{equ:Inplus1}
\end{equation}
where in the second equality Eqs.~(\ref{equ:Indiff}) and (\ref{equ:Inrec}) were 
used. The asymptotically dominant terms in the sums in Eq.~(\ref{equ:cord}) satisfy
$p^2/\dt = \mathcal{O}(1)$. We then note that from the scaling Eq.~(\ref{equ:Inxnlg}) 
one has $\partial_{\dt} I_p(\dt) = I_p(\dt) \left[ 1+\mathcal{O}(1/\dt) \right]$ and, 
via Eq.~(\ref{equ:Inplus1}), $I_{p+1}(\dt) = I_p(\dt) \left[ 1 - p/\dt + 
\mathcal{O}(1/\dt) \right]$. From this the leading terms in the Green's functions are 
\begin{eqnarray}
  \G{(0,1),(-p,q+1)}(\dt) & \sim & \frac{p+q}{\dt} \, \rme^{-2\dt} I_p(\dt) I_q(\dt), 
  \\
  \G{(0,-1),(p,q)}(\dt)   & \sim & \frac{p-q}{\dt} \, \rme^{-2\dt} I_p(\dt) I_q(\dt). 
\end{eqnarray}
The aging expansion of $\cord(t,\tw)$ follows by substituting this into Eq.~(\ref{equ:cord}) 
and using the scalings Eqs.~(\ref{equ:Inxlg}), (\ref{equ:Inxnlg}) and (\ref{equ:Hnxnlg}) for 
the functions $I_n$ and $H_n$. The summations in Eq.~(\ref{equ:cord}) then take the form 
$(1/\tw) \sum_{p,q \geq 0} f\left(p/\sqrt{\tw},q/\sqrt{\tw}\right)$ and become integrals 
in the aging limit,
\begin{eqnarray} 
\fl \cord(t,\tw) \sim \frac{4}{\pi^{3/2}} \frac{\tw}{\dt^2} \int_0^\infty \rmd x \int_0^\infty \rmd y \, 
  (x+y) \, \rme^{-(2\tw/\dt) \left( x^2 + y^2 \right)} \Erfc(x+y) 
  \nonumber \\ 
  - \frac{8}{\pi^{3/2}} \frac{\tw}{\dt^2} \int_0^\infty \rmd x \int_0^\infty \rmd y \, 
  (x-y) \, \rme^{-(2\tw/\dt) \left( x^2 + y^2 \right)} \rme^{-x^2} \Erfc(y)
  \label{equ:cordagint} 
\end{eqnarray}
with $\Erfc(\cdot)$ the complementary error function as defined in
Eq.~(\ref{equ:Hnxnlg}). The integrals in this expression have elementary 
solutions \cite{Mathbook}. One shows that 
\begin{eqnarray}
\fl \cord(t,\tw) \sim \frac{1}{\pi} \frac{1}{t+\tw} \left[ \sqrt{\frac{t+\tw}{t-\tw} } - 1 
  \right. 
  \nonumber \\
  \left. {}+ \frac{2}{\pi} \sqrt{\frac{t-\tw}{2\tw}} \arccos \sqrt{\frac{t-\tw}{t+\tw}} 
  - \frac{1}{\pi} \frac{t+\tw}{\sqrt{t \, \tw}} \arccos \frac{t-\tw}{t+\tw} \right]. 
  \label{equ:cordag}
\end{eqnarray}
This is the exact scaling of $\cord(t,\tw)$ in the aging limit $\dt,\tw \to \infty$ at 
fixed and finite $\dt/\tw$. As emphasised above, it applies not only
for free fermion DLPC but any diffusion coagulation  
process with finite coagulation rate, including in particular the FA 
model. The result Eq.~(\ref{equ:cordag}) takes a simple form in the limiting 
cases 
\begin{eqnarray}
  \dt \ll \tw: \, \cord(t,\tw) \approx \frac{1}{\pi \, \sqrt{2 \, \dt \, \tw}}, 
  \label{equ:corddtlltw} \\ 
  \dt \gg \tw: \, \cord(t,\tw) \approx \frac{3\pi-8}{3\pi^2} \frac{\tw}{\dt^2}. 
  \label{equ:corddtggtw}
\end{eqnarray}
As expected, the $\dt\ll\tw$ limit Eq.~(\ref{equ:corddtlltw}) matches
the expansion for $\dt \gg 1$ of the quasi-equilibrium result 
Eq.~(\ref{equ:cordst}). Plots of the scaling function (\ref{equ:cordag}) 
are shown in Fig.~\ref{fig:d} below and illustrate the crossover 
from the scaling (\ref{equ:corddtlltw}) to (\ref{equ:corddtggtw}). 

It is interesting to compare the above scaling of $\cord(t,\tw)$ in DLPC 
to the cor\-res\-ponding result $C_{\mathrm{d},A}(t,\tw)$ for DLPA. The latter was derived in 
\cite{MayBerGarSol03} and reads~\footnote{Note that $C_{\mathrm{d}}(t,\tw)$ 
in \cite{MayBerGarSol03} is defined in terms of the observable $A = \sigma_i^z 
\sigma_{i+1}^z$ and thus differs from the present choice $A = \n_i = 
\frac{1}{2}(1-\sigma_i^z \sigma_{i+1}^z)$ by a factor $4$; the same applies 
for response functions.} 
\begin{equation}
  C_{\mathrm{d},A}(t,\tw) \sim \frac{1}{2\pi} \frac{1}{\sqrt{t+\tw}}\left[
  \frac{1}{\sqrt{t-\tw}}-\frac{1}{\sqrt{t+\tw}}\right].
  \label{equ:cordAag} 
\end{equation}
Plots of this are also displayed in Fig.~\ref{fig:d} below and demonstrate that the defect autocorrelation 
functions in DLPC and DLPA are qualitatively very similar. Quanti\-tatively, the limit forms 
of Eq.~(\ref{equ:cordAag}) are 
\begin{eqnarray}
  \dt \ll \tw: \, C_{\mathrm{d},A}(t,\tw) \approx \frac{1}{2\pi \sqrt{2 \, \dt \, \tw}},  
  \label{equ:cordAdtlltw} \\ 
  \dt \gg \tw: \, C_{\mathrm{d},A}(t,\tw) \approx \frac{1}{2\pi} \frac{\tw}{\dt^2}. 
  \label{equ:cordAdtggtw}
\end{eqnarray}
From Eqs.~(\ref{equ:corddtlltw},\ref{equ:cordAdtlltw}) we observe that  
$C_{\mathrm{d},A}(t,\tw) \approx \frac{1}{2} \cord(t,\tw)$ in the regime $\dt \ll \tw$. 
This is a direct consequence of the quasi-equilibrium behaviour 
Eq.~(\ref{equ:cordst}), which applies for DLPC as well as DLPA, and 
the fact that defect concentrations are related by $\langle \n_i(\tw) \rangle_a 
= \frac{1}{2} \langle \n_i(\tw) \rangle_c$, c.f.\ Eq.~(\ref{equ:mapn}). 
But in the opposite regime $\dt \gg \tw$ one finds $C_{\mathrm{d},A}(t,\tw) 
\approx r \, \cord(t,\tw)$ with $r = 3\pi/(6\pi-16) \approx  3.31$ from 
Eqs.~(\ref{equ:corddtggtw},\ref{equ:cordAdtggtw}). Here defects interact 
and we thus expect that the nontrivial ratio $r$ is determined by genuine 
many body effects. These must clearly differ between DLPC and DLPA, given that --
amongst other things -- the two processes
have different scaling forms for their domain size distributions \cite{DerZei96}.

\subsection{Local Response Function}

According to Eq.~(\ref{equ:res2tas}) the two-time defect autoresponse function 
in the FA model may be decomposed into two contributions $\resd(t,\tw)=\rad(t,\tw)+\rsd(t,\tw)$. 
As the asymmetric and symmetric perturbations have very different effects on the 
local defect dynamics we discuss the corresponding response functions separately.

\subsubsection{Asymmetric Perturbation}
\label{sec:rad}

The two-time defect autoresponse to the asymmetric part of the perturbation is given by 
Eq.~(\ref{equ:res2ta}) with $n=0$, 
\begin{equation}
  \rad(t,\tw) = \partial_{\tw} \rme^{-2t} I_0(t-\tw) [ I_0 + I_1 ](t+\tw) + \Da_0(t,\tw). 
  \label{equ:resda} 
\end{equation}
The correction $\Da_0(t,\tw)$ is subdominant at large $\tw$ and for any $\dt \geq 0$ 
and will therefore be neglected in the following. Due to the explicit $\tw$-derivative 
in Eq.~(\ref{equ:resda}) the corresponding 
step response function can be obtained directly as
\begin{equation}
\fl \xad(t,\tw) = \int_{\tw}^t \rmd \tau \rad(t,\tau) \sim \rme^{-2t} \left\{ [ I_0 + I_1 ](2t) - 
  I_0(t-\tw) [ I_0 + I_1 ](t+\tw) \right\}. 
  \label{equ:chida} 
\end{equation}
Plots of $\xad(t,\tw)$ are shown in Fig.~\ref{fig:d} below. This result is interesting 
in various ways. Consider first the quasi-equilibrium behaviour at fixed $\dt \geq 0$ 
and large $\tw$, 
\begin{equation}
  \xad(t,\tw) \sim \frac{1}{\sqrt{\pi \tw}} \left[ 1 - \rme^{-\dt} I_0(\dt) \right] 
  \sim c(\tw) \left[ 1 - p_r(\dt) \right], 
  \label{equ:chidast} 
\end{equation}
to leading order in $\tw$. The form of Eq.~(\ref{equ:chidast}) reflects a simple 
mechanism: at time $\tw$ there is a homogeneous concentration $c(\tw)$ of defects in 
the system. But application of 
the asymmetric perturbation $\Va_i$, which traps diffusing defects on site $i$, 
yields an increase of $\langle \n_i(t) \rangle$ as given by Eq.~(\ref{equ:chidast}). 
The step response $\xad(t,\tw)$ approaches $c(\tw)$ within a time interval 
$\dt = \mathcal{O}(1)$, see Fig.~\ref{fig:d}. In the aging regime, where both 
$\dt$ and $\tw$ are large and comparable, one has from Eq.~(\ref{equ:chida}) 
that $\xad(t,\tw) \sim c(t)$. The step response then decreases along with the 
density of defects in the system. This explains the non-monotonic shape of 
$\xad(t,\tw)$ shown in Fig.~\ref{fig:d}. 
\begin{figure}[t]
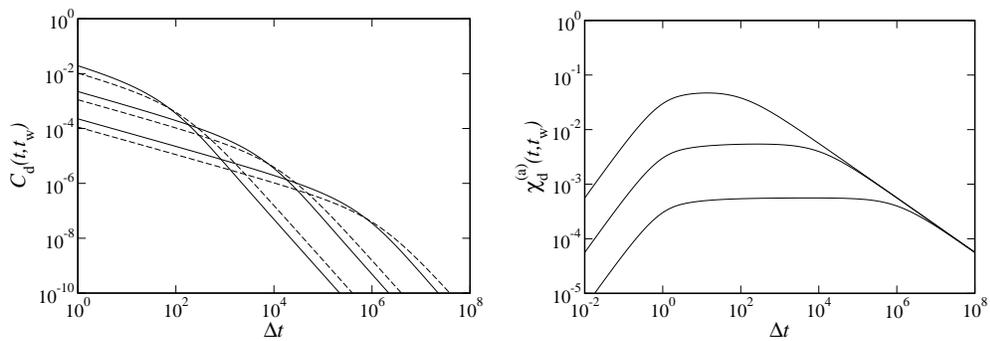

	\hspace*{1in} 
	\epsfig{file=Cd.eps, width=6.3cm, clip}
	\hfill
	\epsfig{file=Xd.eps, width=6.3cm, clip} 
  \caption{\label{fig:d} Left panel: scaling form of the two-time defect autocorrelation 
    $\cord(t,\tw)$, Eq.~(\ref{equ:cordag}), in the FA model for $\tw = 10^2, 10^4$ and $10^6$ (full 
    lines) and, for comparison, the corresponding DLPA result Eq.~(\ref{equ:cordAag}) 
    at the same values of $\tw$ (dashed lines). Right panel: plots of the step response 
    $\xad(t,\tw)$ to the asymmetric part of the perturbation, Eq.~(\ref{equ:chida}), for $\tw = 10^2, 10^4$ 
    and $10^6$ from top to bottom, respectively. 
  }
\end{figure}

It is important to note that $\xad(t,\tw)$ is dominated by the small-$\dt$ 
behaviour of $\rad(t,\tw)$. Indeed, in the quasi-equilibrium regime 
Eq.~(\ref{equ:chidast}) reproduces 
\begin{equation} 
  \rad(t,\tw) = -\partial_{\tw} \xad(t,\tw) \sim - c(\tw) \, \partial_{\dt} p_r(\dt), 
  \label{equ:resdadtO1}
\end{equation}
consistent with Eq.~(\ref{equ:resda}); this result is positive since 
$p_r(\dt)$ is a decreasing function of $\dt$ and hence $-\partial_{\dt} p_r(\dt) > 0$. 
But in the aging regime $\xad(t,\tw) \sim c(t)$ to leading order as discussed above, 
which would naively suggest $\rad(t,\tw) = -\partial_{\tw} c(t) = 0$. More precisely it means that the aging part of 
$\rad(t,\tw)$ only contributes subdominantly to $\xad(t,\tw)$. From the full expression 
Eq.~(\ref{equ:chida}), or equivalently Eq.~(\ref{equ:resda}), one deduces that in fact 
\begin{equation}
  \rad(t,\tw) \sim \frac{1}{2\pi} \left[ \frac{1}{\sqrt{t+\tw} (t-\tw)^{3/2}} - 
  \frac{1}{\sqrt{t-\tw} (t+\tw)^{3/2}} \right]. 
  \label{equ:resdaag}
\end{equation}
For $\dt \ll \tw$ the first term in Eq.~(\ref{equ:resdaag}) dominates, matching 
the large $\dt$ behaviour of the quasi-equilibrium result Eq.~(\ref{equ:resdadtO1}). 
For $\dt \gg \tw$, on the other hand, the terms in Eq.~(\ref{equ:resdaag}) cancel to 
leading order in $\tw/\dt$; then $\rad(t,\tw) \approx \tw/(\pi\,\dt^3)$. 

We add that a response function analogous to $\rad(t,\tw)$ was calculated 
for DLPA in \cite{MayBerGarSol03}. Denoting the latter by $R_{\mathrm{d},A}(t,\tw)$ 
one has from \cite{MayBerGarSol03} that $R_{\mathrm{d},A}(t,\tw) = \frac{1}{2} \rad(t,\tw)$. 
In terms of the two-time defect autoresponse for an asymmetric perturbation there is therefore no distinction
between DLPC and DLPA, except for the factor of $\frac{1}{2}$ 
which has its origin in $\langle \n_i(\tw) \rangle_a = \frac{1}{2} \langle \n_i(\tw) \rangle_c$, 
see Eq.~(\ref{equ:mapn}).

\subsubsection{Symmetric Perturbation}

Setting $n=0$ in Eq.~(\ref{equ:res2ts}) gives for the autoresponse to the symmetric part of 
the perturbation, 
\begin{equation}
  \rsd(t,\tw) = - \frac{1}{2} \rme^{-2t} I_0(t-\tw) [ I_0 - I_2 ](t+\tw) + \Ds_0(t,\tw). 
  \label{equ:resds} 
\end{equation} 
As discussed in Sec.~\ref{sec:res2ts} the behaviour for short time
differences $\dt = \mathcal{O}(1)$ 
of this result does depend on the precise coagulation rate. So in this regime 
Eq.~(\ref{equ:resds}) applies to free fermion DLPC but not the
FA model. This does  
not constitute a problem, however, since Eq.~(\ref{equ:resds}) scales as
$\mathcal{O}\left({\tw}^{-3/2}\right)$ at large $\tw$ and fixed $\dt$. It is therefore 
subdominant against the short-time scaling of the asymmetric response 
Eq.~(\ref{equ:resdadtO1}). 
Only in the aging regime does the
symmetric response contribute significantly to the overall
response function; there it is independent of the coagulation rate 
as explained above and so does apply to the FA model. 
Taking the appropriate limits in Eq.~(\ref{equ:resds}) gives 
\begin{equation}
  \rsd(t,\tw) \sim -\frac{1}{2\pi} \frac{1}{\sqrt{t-\tw} \, (t+\tw)^{3/2}}. 
  \label{equ:resdsag}
\end{equation}
For $\dt \ll \tw$ this simplifies to $\rsd(t,\tw) \approx 
- 1/\big[2\pi \sqrt{\dt} (2\tw)^{3/2}\big]$ 
while for $\dt \gg \tw$ one has $\rsd(t,\tw) \approx
-1/(2\pi \dt^2)$. Note the overall {\em negative} sign of the response: 
to understand this result qualitatively recall that the symmetric 
perturbation $\Vs_i$ increases the local diffusion rates between 
sites $i-1 \leftrightharpoons i \leftrightharpoons i+1$. A response 
can only occur if a defect is present on one of these sites at time 
$\tw$. During the time interval where the perturbation is applied 
the dynamics of this defect are accelerated. In and by itself this 
does not have any effect on the defect concentration $\langle \n_i(t) \rangle$  
at site $i$. However, the ``extra time" for the defect manifests itself 
through the fact that it will now coagulate earlier with a neighbouring 
defect. The local concentration is therefore 
reduced, and this explains the negative sign of the response. 
Physically this behaviour derives from the activated character of the FA
dynamics \cite{MayLeoBerGarSol06}. The increased local diffusion rates 
in the effective FA model represent locally reduced energy barriers 
for the creation of defects. In equilibrium this would increase the 
defect density; out of equilibrium the dominant effect is a speeding up 
of the relaxation of the defect density towards zero, leading to a
reduction in the number of defects.

The scaling of the step response function $\xsd(t,\tw)$ in the aging 
regime is obtained correctly by integration of Eq.~(\ref{equ:resdsag}), 
the reason being that small-$\dt$ contributions from $\rsd(t,\tw)$ 
only amount to subdominant corrections in the integral. Thus, 
\begin{equation}
  \xsd(t,\tw)  = \int_{\tw}^t \rmd \tau \rsd(t,\tau) \sim - \frac{1}{2\pi} \frac{\sqrt{t-\tw}}{t \sqrt{t+\tw}}, 
  \label{equ:chidsag}
\end{equation}
in the aging regime $\dt,\tw \gg 1$ with fixed and finite $\dt/\tw$. 
For $\dt \ll \tw$ this reduces to $\xsd(t,\tw) \approx 
-\sqrt{\dt}/(2\sqrt{2}\pi {\tw}^{3/2})$ while for $\dt \gg \tw$ one 
finds $\xsd(t,\tw) \approx -1/(2\pi \dt)$. Comparison of Eq.~(\ref{equ:chidsag}) 
with Eq.~(\ref{equ:chida}) shows that the overall step response 
in the aging regime $\chi_{\mathrm{d}}(t,\tw) = \xad(t,\tw) + \xsd(t,\tw) 
\sim c(t)$ is dominated by the short-$\dt$ behaviour of $\rad(t,\tw)$ as 
discussed above. The aging parts of $\rad(t,\tw)$, Eq.~(\ref{equ:resdaag}), 
and $\rsd(t,\tw)$, Eq.~(\ref{equ:resdsag}), are both of the same order 
but only contribute subdominantly in the overall step response 
$\chi_\mathrm{d}(t,\tw)$. The consequences of this will become clear 
below.

\subsection{FDR and FD plot}
\label{sec:dooda}

Based on the above results for connected two-time defect autocorrelation and 
response functions one immediately obtains the associated fluctuation-dissipation
(FD) plots. We use normalized FD plots \cite{MayBerGarSol03,SolFieMay02,MayBerGarSol04} 
of $\widetilde{\chi}(t,\tw) = \chi(t,\tw)/C(t,t)$ versus $1 - \widetilde{C}(t,\tw)$ 
where $\widetilde{C}(t,\tw) = C(t,\tw)/C(t,t)$. Normalization scales the plot 
so that the abscissa always lies in the range from zero to one. It is crucial 
that the plot is parameterised by $\tw$ with $t$ held fixed~\cite{FieSol02,SolFieMay02}. Only then is the 
slope of the plot guaranteed to be $X(t,\tw)$ as one readily verifies 
from Eq.~(\ref{equ:gFDT}) and $R(t,\tw) = - \partial_{\tw} \chi(t,\tw)$. 

\begin{figure}[t]
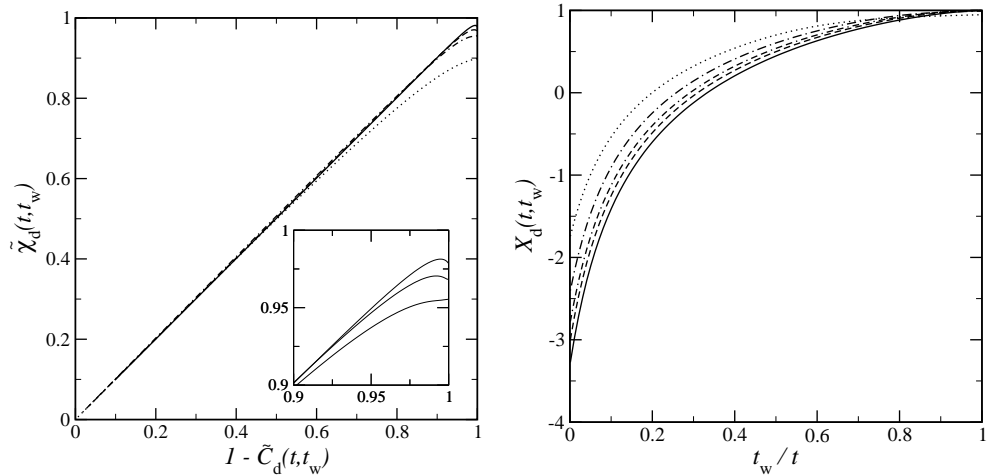

  \hspace*{1in} 
  \epsfig{file=FD-Plot-Local-t.eps, width=6.3cm, clip}
  \hfill
  \epsfig{file=FDR-Local-tw.eps, width=6.3cm, clip}
  \caption{\label{fig:FDlocal} Left: Normalised FD plot for the local defect 
  observable $A=\n_i$. Data were obtained 
  by numerical evaluation \cite{Thesis} of the full expressions (\ref{equ:cord}),
  (\ref{equ:resda}), (\ref{equ:resds}) 
  at $t=10^{1/2},10^1,10^{3/2},10^2$ with $\tw \in [0,t]$. For increasing $t$ 
  the curves move upwards, converging to the equilibrium line of unit
  slope. Inset: The data in the top-right 
  corner on a magnified scale, not containing the $t=10^{1/2}$ curve. Right: 
  Plot of the FDR $\Xd(t,\tw)$ versus $\tw/t$ for $\tw=10^{1/2}$ (dotted), 
  $10^1,10^{3/2},10^2$ and $\infty$ (full line). The curves for finite 
  $\tw$ were again obtained by numerical evaluation of the full expressions while 
  the limit curve follows from the scaling forms (\ref{equ:cordag}), 
  (\ref{equ:resdaag}) and (\ref{equ:resdsag}).}
\end{figure}

The FD plot for the local defect observable in the FA model is shown in 
Fig.~\ref{fig:FDlocal}. It quickly converges 
to the equilibrium line $\widetilde{\chi}_\mathrm{d} = 1 - \widetilde{C}_\mathrm{d}$, 
consistently with observations made in computer simulations \cite{BuhGar02,Buh03}. 
However, a plot of the FDR $\Xd(t,\tw)$, also shown in Fig.~\ref{fig:FDlocal}, clearly 
demonstrates that equilibrium FDT is {\em not} satisfied. In fact $\Xd(t,\tw)$ 
steadily decreases with increasing $t$ and actually turns {\em negative} 
around $\tw/t \approx 1/3$. This is reflected in the FD plot, which bends down 
as $\widetilde{C}_\mathrm{d} \to 0$. The plot of $\Xd(t,\tw)$ also indicates that 
a nontrivial scaling form $\Xd(t,\tw) \sim \Xd(\tw/t)$ exists in the
aging limit; we return to this below.

In order to understand the shape of the FD plot let us first consider the 
dynamics in the quasi-equilibrium regime of large $\tw$ but finite $\dt \geq 0$. 
Here $C_\mathrm{d}(t,\tw) \sim c(\tw) p_r(\dt)$ as given by Eq.~(\ref{equ:cordst}) 
and $\chi_\mathrm{d}(t,\tw) \sim \xad(t,\tw) \sim c(\tw) [ 1 - p_r(\dt)]$, 
Eq.~(\ref{equ:chidast}), is dominated by the asymmetric 
perturbation. To leading order in $\tw$ we then have for the normalized 
quantities $\widetilde{\chi}_\mathrm{d} = 1 - \widetilde{C}_\mathrm{d} = 
1-p_r(\dt)$ and equilibrium FDT is satisfied. Furthermore, since 
$\widetilde{C}_\mathrm{d} \sim p_r(\dt)$ becomes negligibly small already 
in the $\dt = \mathcal{O}(1)$ time sector, aging effects are compressed into the 
top right corner of the FD plot. Asymptotically only the 
quasi-equilibrium regime is visible in the FD plot 
\cite{MayBerGarSol03,MaySol05,MayLeoBerGarSol06}. 

Consider next the aging regime where both $\dt$ and $\tw$ are large, but 
with a finite ratio $\dt/\tw$. The normalized correlation then is 
$\widetilde{C}_\mathrm{d} = \mathcal{O}(1/\sqrt{t})$. The dominant 
scaling of the step response $\chi_\mathrm{d} \sim c(t)$ would give
$\widetilde{\chi}_\mathrm{d} = 1$. This is modified by aging contributions 
from $\rad(t,\tw)$ and $\rsd(t,\tw)$ so that $1-\widetilde{\chi}_\mathrm{d} 
= \mathcal{O}(1/\sqrt{t})$. In the FD plot, aging effects are only visible 
in a region of size $\mathcal{O}(1/\sqrt{t})$ in the top right corner of 
the plot; this regime is highlighted in the inset of Fig.~\ref{fig:FDlocal}. 
Effects of the asymmetric perturbation alone, which are accounted for 
in $\xad(t,\tw)$, Eq.~(\ref{equ:chida}), would cause the plot to turn 
horizontal with $X^\infty = 0$; this would be the analog of the DLPA 
result in \cite{MayBerGarSol03}. However, the additional {\em negative} 
contribution of $\xsd(t,\tw)$, Eq.~(\ref{equ:chidsag}), in the overall 
step response $\chi_\mathrm{d}(t,\tw) = \xad(t,\tw) + \xsd(t,\tw)$ 
makes the plot non-monotonic with $X^\infty < 0$, which is 
consistent with simulations \cite{MayLeoBerGarSol06}. Note that this 
non-monotonicity is genuine and {\em not} due to incorrect parameterisation 
of the plot with $t$ instead of $\tw$ \cite{CriRitRocSel00}.

The above discussion shows that for defect observables an FD plot gives a rather
unsatisfactory graphical representation 
of the FDT violation in the FA model. It is more appropriate to 
look instead directly at the time dependence of the FDR $\Xd(t,\tw)$, as shown 
in Fig.~\ref{fig:FDlocal}. The FDR involves the response functions 
$\rad(t,\tw)$ and $\rsd(t,\tw)$. In contrast to the associated step 
response functions, these are not dominated only by the quasi-equilbrium
dynamics at small $\dt$ and so can reveal genuine aging effects. Looking at
the relative importance of the asymmetric and symmetric contributions, one sees that
in the aging regime $\dt,\tw \gg 1$ with $\dt/\tw = \mathcal{O}(1)$ 
both $\rad(t,\tw)$, Eq.~(\ref{equ:resdaag}), and $\rsd(t,\tw)$, 
Eq.~(\ref{equ:resdsag}), are of the same order $\mathcal{O}(1/{\tw}^2)$. 
Keeping in mind that $\rad(t,\tw) > 0$ while $\rsd(t,\tw) < 0$ and looking 
at the plot of $\Xd(t,\tw)$ in Figure~\ref{fig:FDlocal}, we notice that
$\rsd(t,\tw)$ in fact becomes dominant for $\dt \gg \tw$; explicitly, 
$\rsd(t,\tw) \approx -1/(2\pi\dt^2)$ while $\rad(t,\tw) \approx \tw/(\pi \dt^3) 
= \mathcal{O}\big((\tw/\dt) |\rsd(t,\tw)|\big)$. This causes the change 
of sign in $\resd(t,\tw)$ and thus $\Xd(t,\tw)$. So, while in the 
$\dt \ll \tw$ limit the response is dominated by the trapping effect 
of the asymmetric part of the perturbation, it is the speedup effect of the local 
dynamics through the symmetric component of the perturbation that dominates for $\dt \gg \tw$. 
This speedup is the consequence of coupling of the observable to 
energy barriers in the activated dynamics of the FA model. 

By working out the scaling of $\partial_{\tw} \cord(t,\tw)$ from 
Eq.~(\ref{equ:cordag}) and using the results Eqs.~(\ref{equ:resdaag},
\ref{equ:resdsag}) one obtains the scaling form of $\Xd(t,\tw)$. 
We do not state the explicit result since it is somewhat bulky; 
importantly, however, its time dependence is only via the ratio 
$\tw/t$. A plot of $\Xd(\tw/t)$ is contained in Figure~\ref{fig:FDlocal}. 
It crosses over from $\Xd=1$ in the quasi-equilibrium limit to $\Xd=\Xd^\infty$ 
for $t\gg\tw$; the asymptotic FDR is 
\begin{equation}
  X_{\mathrm{d}}^\infty = -\frac{3 \, \pi}{6 \, \pi - 16} \approx -3.307.
  \label{equ:Xdinfty}
\end{equation}
We recall that the results shown in Fig.~\ref{fig:FDlocal} are fully
exact for the free fermion DLPC process (with filled initial state).
In the limit of large $\tw$, where the precise coagulation rate and
initial state density become irrelevant, the results -- including the scaling form of $\Xd(\tw/t)$ 
and the value $\Xd^\infty$ -- apply to
arbitrary DLPC processes and, in particular, to the FA model.
Finite time data were included in 
Figure~\ref{fig:FDlocal} only in order to given an indication of 
the speed of convergence (for free fermion DLPC) to the long-time 
asymptotic results.

\section{Global Correlation and Response}
\label{sec:global_fdt}

Given that the FA model can be viewed as a coarsening system, and following the philosophy 
of \cite{MayBerGarSol03}, we now analyse the nonequilibrium FDT for the 
global observable $A=\sum_i \n_i$, which is nothing but the energy. Our 
expectations are, firstly, that we should find the same $X^\infty$ as for the 
local defect observable \cite{MayBerGarSol03,CalGam05} and, secondly,
that a nontrivial limit FD plot should exist in this case \cite{MayBerGarSol03,
MayBerGarSol04,Pleimling04}. 
As above we first study separately the connected two-time energy correlation 
and response functions $\core(t,\tw)$ and $\rese(t,\tw)$,
respectively; both will be normalised by $1/N$ to get quantities of order unity.
Then the resulting FDR and FD plot are discussed.

\subsection{Energy Correlation}
\label{sec:core}

We obtain an expression for $\core(t,\tw)$ by summing Eq.~(\ref{equ:cor2tsums}) over $n$ 
since, using translational invariance,  $\core(t,\tw) = (1/N) \sum_{i,j}[\langle 
\n_i(t) \n_j(\tw) \rangle - \langle \n_i(t) \rangle \langle 
\n_j(\tw) \rangle] = \sum_n C_n(t,\tw)$. Via the convolution property Eq.~(\ref{equ:Inconv}) 
one readily shows that  
\begin{eqnarray}
  \sum_n \G{(n,n+1),(-p,q+q)}(\dt) & = & \rme^{-2\dt} [I_{p+q}- I_{p+q+2}](2\dt), \\ 
  \sum_n \G{(\pm n ,\pm n -1),(p,q)}(\dt) & = & \rme^{-2\dt} [I_{p-q-1} - I_{p-q+1}](2\dt). 
\end{eqnarray}
Thus we have from Eq.~(\ref{equ:cor2tsums})  
\begin{eqnarray}
\fl \core(t,\tw) = 
	\rme^{-2t} [I_0 + I_1](2\tw) \sum_{p,q=0}^\infty [I_{p+q}- I_{p+q+2}](2\dt) 
	\, H_{p+q+1}(2\tw) 
  \nonumber \\ 
  - 2 \sum_{p,q=0}^\infty \rme^{-2t} [I_{p-q-1} - I_{p-q+1}](2\dt) \, [I_p + I_{p+1}](2\tw) 
  \left[ \delta_{q,0} + H_q(2\tw) \right]. 
  \label{equ:core}
\end{eqnarray}
The energy correlation function $\core(t,\tw)$ is constant -- with respect 
to $\dt$ -- in the quasi-equilibrium regime. This is obvious when considering 
that defects are typically a distance $\mathcal{O}\left(\sqrt{\tw}\right)$ apart and can 
therefore not meet within a time interval $\dt=\mathcal{O}(1)$. Thus no coagulation 
processes are possible and the energy remains constant. 
It is somewhat awkward to derive this directly from
Eq.~(\ref{equ:core}) because the double sum in the second line only
reduces to a one-dimensional sum; the assumption $\dt = \mathcal{O}(1)$ 
imposes merely that $p-q$ must be of $\mathcal{O}(1)$ to get a significant contribution. One can get
the result indirectly, however, by first going to the aging regime 
and then considering the limit $\dt\ll\tw$ (see below). The
correlation function at $\dt=0$ itself, which gives the energy
fluctuations, is much simpler to work out using $I_n(0) = \delta_{n,0}$. 
In this case it can be shown that 
\begin{equation}
  \core(\tw,\tw) = 3 \rme^{-2\tw} [I_0 + I_1](2\tw) - \rme^{-4\tw} [3 I_0 + 4 I_1 + I_2](4\tw) 
  \label{equ:coretwtw} 
\end{equation}
An aging expansion of Eq.~(\ref{equ:core}) is obtained rather easily using Eqs.~(\ref{equ:Inrec}), 
(\ref{equ:Inxlg}), (\ref{equ:Inxnlg}) and (\ref{equ:Hnxnlg}). In the aging limit the 
double sums turn into integrals and one finds 
\begin{eqnarray}
\fl \core(t,\tw) \sim \frac{4}{\pi} \frac{\tw}{\dt^{3/2}} \int_0^\infty \rmd x \int_0^\infty \rmd y \, 
  (x+y) \rme^{-(\tw/\dt) (x+y)^2} \Erfc(x+y) 
  \nonumber \\ 
  - \frac{8}{\pi} \frac{\tw}{\dt^{3/2}} \int_0^\infty \rmd x \int_0^\infty \rmd y \, 
  (x-y) \rme^{-(\tw/\dt) (x-y)^2} \rme^{-x^2} \Erfc(y). 
  \label{equ:coreint}
\end{eqnarray}
Again there are elementary solutions to these integrals \cite{Mathbook}, giving 
\begin{eqnarray}
\fl \core(t,\tw) \sim \frac{2}{\sqrt{\pi}} \left[ 
  \frac{1}{\sqrt{t}} - \frac{1}{\sqrt{t+\tw}} - \frac{\sqrt{t-\tw}}{\pi \, t} \right]
  \nonumber \\
  +\frac{2}{\pi^{3/2}} \left[ \frac{1}{\sqrt{\tw}} \arcsin
  \sqrt{\frac{\tw}{t}} - \frac{2}{\sqrt{t+\tw}} \arcsin 
  \frac{\tw}{t} \right]. 
  \label{equ:coreag}
\end{eqnarray}
This leading order long-time scaling applies to all DLPC
processes with finite coagulation rate 
in the limit $\tw \gg 1$ and for any $\dt \geq 0$. 
Coagulation rate effects are in principle present for
$\dt=\mathcal{O}(1)$ but are subleading for large $\tw$ because they 
are driven by the density of small domains.
For $\dt \ll \tw$ the expression~(\ref{equ:coreag}) approaches a
constant value as anticipated and matches up with the 
large-$\tw$ expansion of the equal-time value, Eq.~(\ref{equ:coretwtw}). 
Specifically, the limit forms of Eq.~(\ref{equ:coreag}) are, 
\begin{eqnarray}
  \dt \ll \tw: \, \core(t,\tw) \approx \frac{3-2\sqrt{2}}{\sqrt{\pi \, \tw}}, 
  \label{equ:coreagdtlltw} \\ 
  \dt \gg \tw: \, \core(t,\tw) \approx \frac{3\pi-8}{3\pi^{3/2}} \frac{\tw}{\dt^{3/2}}. 
  \label{equ:coreagdtggtw}
\end{eqnarray}
Figure~\ref{fig:e} below shows plots of the scaling function Eq.~(\ref{equ:coreag}), 
illustrating the crossover from the plateau Eq.~(\ref{equ:coreagdtlltw}) to the long 
time scaling Eq.~(\ref{equ:coreagdtggtw}). 

In analogy with the discussion in Sec.~\ref{sec:cord} we again compare the 
scaling Eq.~(\ref{equ:coreag}) of $\core(t,\tw)$ in DLPC with the corresponding result 
$C_{\mathrm{e},A}(t,\tw)$ in DLPA. From Ref.~\cite{MayBerGarSol03}, 
\begin{equation}
  C_{\mathrm{e},A}(t,\tw) \sim \frac{1}{\sqrt{\pi}} \left( \frac{1}{\sqrt{t}} - \frac{1}{\sqrt{t+\tw}}\right), 
  \label{equ:coreAag}
\end{equation}
bearing in mind that the result in \cite{MayBerGarSol03} contains an extra factor $4$. 
Plots of Eq.~(\ref{equ:coreAag}) are also included in Fig.~\ref{fig:e} below, and 
cross over between the limits 
\begin{eqnarray}
  \dt \ll \tw: \, C_{\mathrm{e},A}(t,\tw) \approx \frac{2-\sqrt{2}}{2\sqrt{\pi \tw}}, 
  \label{equ:coreAagdtlltw} \\ 
  \dt \gg \tw: \, C_{\mathrm{e},A}(t,\tw) \approx \frac{1}{2\sqrt{\pi}} \frac{\tw}{\dt^{3/2}}. 
  \label{equ:coreAagdtggtw} 
\end{eqnarray}
In the regime $\dt \ll \tw$ we can again establish a simple relationship between 
the functions $C_{\mathrm{e},A}(t,\tw)$ and $\core(t,\tw)$. It is sufficient to consider 
the instantaneous energy fluctuations for $t=\tw$ since both functions are, to leading 
order, independent of $\dt$ when $\dt \ll \tw$. Now, by definition $C_{\mathrm{e},A}(\tw,\tw)
= (1/N) \sum_{i,j} [\langle \n_i(\tw) \n_j(\tw) \rangle_a - \langle \n_i(\tw) \rangle_a 
\langle \n_j(\tw) \rangle_a]$. Using the mapping Eq.~(\ref{equ:mapn}) we have, for 
$i \neq j$, $\langle \n_i(\tw) \n_j(\tw) \rangle_a = \frac{1}{4} \langle \n_i(\tw) \n_j(\tw) \rangle_c$. 
However, the diagonal contributions $i = j$ in this sum yield $\langle \n_i(\tw) \n_i(\tw) \rangle_a = 
\langle \n_i(\tw) \rangle_a = \frac{1}{2} \langle \n_i(\tw) \rangle_c$. Keeping this 
in mind one readily shows that $C_{\mathrm{e},A}(\tw,\tw) = \frac{1}{4} \left[ \core(\tw,\tw) 
+ c(\tw) \right]$ with $c(\tw) = \langle \n_i(\tw) \rangle_c \sim 1/\sqrt{\pi \tw}$ as usual. 
This identity is indeed satisfied by Eqs.~(\ref{equ:coreagdtlltw},\ref{equ:coreAagdtlltw}). 
While we found $C_{\mathrm{d},A}(t,\tw) \approx \frac{1}{2} \cord(t,\tw)$ for defect correlations 
in the regime $\dt \ll \tw$, the energy fluctuations are related by $C_{\mathrm{e},A}(t,\tw) \approx 
r_0 \core(t,\tw)$ with $r_0 = 1+1/\sqrt{2} \approx 1.71$. Interestingly, in the opposite regime 
$\dt \gg \tw$ we have from Eqs.~(\ref{equ:coreagdtggtw},\ref{equ:coreAagdtggtw}) that 
$C_{\mathrm{e},A}(t,\tw) \approx r \, \core(t,\tw)$ with the same nontrivial ratio 
$r = 3\pi/(6\pi-16) \approx  3.31$ as for the defect correlations.~\footnote{We note 
that the ratio $r$ coincides with the absolute value of the asymptotic FDR $\Xd^\infty$, 
Eq.~(\ref{equ:Xdinfty}). For defect observables this originates from the fact that 
$\partial_{\tw} C_{\mathrm{d},A}(t,\tw) \sim 1/(2\pi t^2)$ in DLPA for $t \gg \tw$, see 
Eq.~(\ref{equ:cordAdtggtw}), scales in the same way as $\rsd(t,\tw) \sim - 1/(2\pi t^2)$ 
in DLPC from Eq.~(\ref{equ:resdsag}). Likewise for the energy $\partial_{\tw} C_{\mathrm{e},A}(t,\tw) 
\sim 1/(2\sqrt{\pi} t^{3/2})$ in DLPA for $t \gg \tw$ according to Eq.~(\ref{equ:coreAagdtggtw}) 
matches the scaling of $\rse(t,\tw) \sim - 1/(2\sqrt{\pi} t^{3/2})$ in DLPC from Eq.~(\ref{equ:resesag}) 
below. However, this seems to be a mere mathematical coincidence with no obvious physical 
implications.}

\subsection{Energy-Temperature Response}

By similar arguments as for the correlation one sees that the
normalised energy response to a 
field acting uniformly on all sites $i$ of the system is given by
$\rese(t,\tw) = \sum_n R_n(t,\tw)$. Using the decomposition of
$R_n(t,\tw)$, Eq.~(\ref{equ:res2tas}), we again discuss 
the asymmetric, $\rae(t,\tw) = \sum_n \ra_n(t,\tw)$, and symmetric, $\rse(t,\tw) =
\sum_n \rs_n(t,\tw)$, parts separately.

\subsubsection{Asymmetric Perturbation}
\label{sec:rae}

Summing $\ra_n(t,\tw)$, Eq.~(\ref{equ:res2ta}), over $n$ and using the convolution 
property Eq.~(\ref{equ:Inconv}) gives 
\begin{equation}
  \rae(t,\tw) = \partial_{\tw} \rme^{-2t} [I_0+I_1](2t) + \Dea(t,\tw) = 0.
	\label{equ:resea}   
\end{equation}
The first term in Eq.~(\ref{equ:resea}) is independent of $\tw$ and 
hence the derivative vanishes trivially. It can be shown that the subdominant 
correction $\Dea(t,\tw)$ also vanishes exactly, so that altogether $\rae(t,\tw) = 0$. 
With hindsight this result is obvious: when applied to all sites, 
the asymmetric perturbations of the local diffusion rates cancel. 
We remark that the same is true for the analogous response function 
$R_{\mathrm{e},A}(t,\tw)$ in DLPA studied in Ref.~\cite{MayBerGarSol03}.

\subsubsection{Symmetric Perturbation}

\begin{figure}[t]
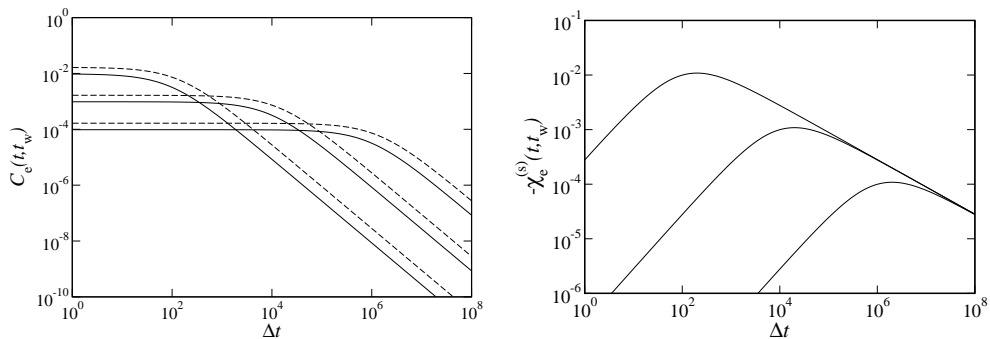

	\hspace*{1in} 
	\epsfig{file=Ce.eps, width=6.3cm, clip}
	\hfill
	\epsfig{file=Xe.eps, width=6.3cm, clip}
  \caption{\label{fig:e} Left panel: scaling form of the normalized two-time energy 
    correlation $\core(t,\tw)$, Eq.~(\ref{equ:coreag}), in the FA model for $\tw = 10^2, 10^4$ and $10^6$ (full 
    lines) and, for comparison, the corresponding DLPA result Eq.~(\ref{equ:coreAag}) 
    at the same values of $\tw$ (dashed lines). Right panel: scaling plots of the energy-temperature 
    step response $-\chi_{\mathrm{e}}(t,\tw) = -\xse(t,\tw)$ obtained from Eq.~(\ref{equ:resesag}) 
    for $\tw = 10^2, 10^4$ and $10^6$ from top to bottom, respectively. 
  }
\end{figure}

The sum over $\rs_n(t,\tw)$, Eq.~(\ref{equ:res2ts}), follows again straightforwardly 
using the convolution property Eq.~(\ref{equ:Inconv}). Here we find 
\begin{equation}
  \rse(t,\tw) = -\rme^{-2t} [I_0 - I_2](2t) + \Des(t,\tw). 
  \label{equ:reses} 
\end{equation} 
The first term is exactly $\partial_t c(t) = -\rme^{-2t} [I_0 - I_2](2t)$ 
with $c(t)$ as given in Eq.~(\ref{equ:cn}). 
This result is easily explained: first, since the
response is normalised by $1/N$ it measures changes in the defect concentration
$c(t) = (1/N) \sum_i \langle \n_i(t) \rangle$, the
latter being just the normalised version of the energy. Now consider
the effect of the corresponding symmetric perturbation. When applied
locally it increases the diffusion rates between sites $i-1
\leftrightharpoons i \leftrightharpoons i+1$. Thus, in the global
case, all diffusion rates are increased and the perturbation
essentially speeds up the dynamics of the system. This is equivalent
to saying that the system gains some extra time $\delta t$ to evolve
and therefore $\rse(t,\tw) = [c(t+\delta t) - c(t)]/\delta t = \partial_t c(t)$. 
In the aging regime this gives 
\begin{equation}
  \rse(t,\tw) \sim  \partial_t c(t) \sim - \frac{1}{2 \, \sqrt{\pi} \, t^{3/2}}. 
  \label{equ:resesag}
\end{equation}
The global symmetric perturbation represents globally reduced energy barriers in the 
activated dynamics of the FA model, thus there is accelerated relaxation 
and a negative response. The scaling $\xse(t,\tw) \sim - 
(t-\tw)/(2 \sqrt{\pi} t^{3/2})$ follows immediately from 
Eq.~(\ref{equ:resesag})
and is shown in Fig.~\ref{fig:e}. It crosses over from 
$\xse(t,\tw) \approx - \dt/(2 \sqrt{\pi} \tw^{3/2})$ 
for $\dt \ll \tw$ to $\xse(t,\tw) \approx - 1/(2 \sqrt{\pi \dt})$ 
when $\dt \gg \tw$. 

The above reasoning applies for $\dt \gg 1$ 
where the second term in Eq.~(\ref{equ:reses}) is subdominant. For
small $\dt$, it cannot be neglected, and accounts for
deviations arising from the fact that the perturbation does not
increase all rates: that is, coagulation rates are left unchanged. Since
coagulation processes are the only ones that can give an instantaneous
decrease of the energy, the instantaneous response must then in fact
vanish, $\rse(\tw,\tw)=0$, and this can be verified from
Eq.~(\ref{equ:reses}) by explicitly evaluating $\Des(\tw,\tw)$.

\subsection{FDR and FD plot}

\begin{figure}[t]
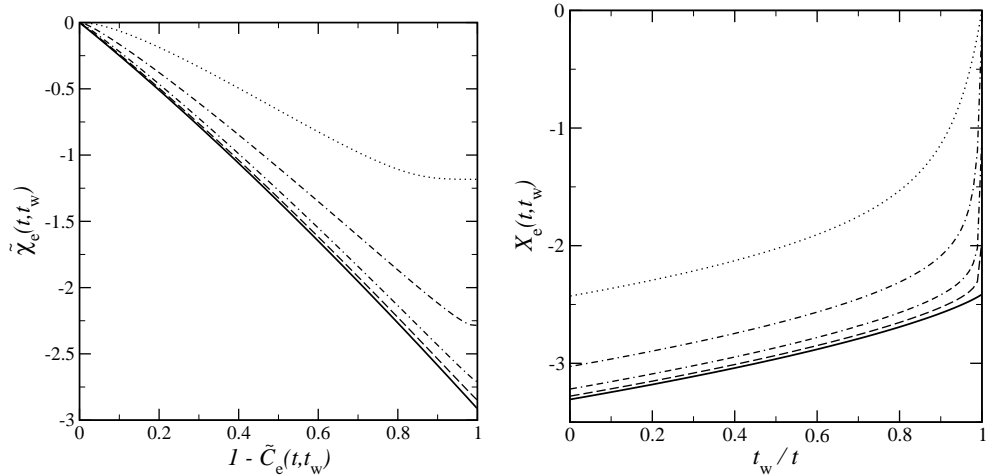

	\hspace*{1in}
  \epsfig{file=FD-Plot-Uniform-t.eps, width=6.3cm, clip}
  \hfill
  \epsfig{file=FDR-Uniform-tw.eps, width=6.3cm, clip}
  \caption{\label{fig:FDglobal} Left: Normalised FD plot for the energy. 
  The curves correspond to $t=10^1$ (dotted), $10^2,10^3,10^4$ and $\infty$ 
  (full curve) with $\tw\in[0,t]$. Finite time data is obtained by numerical 
  evaluation \cite{Thesis} of the full expressions (\ref{equ:core}) and (\ref{equ:reses}) 
  while the limit curve 
  follows from Eqs.~(\ref{equ:coreag},\ref{equ:resesag}). Right: Plots of 
  $\Xe(t,\tw)$ versus $\tw/t$ for $\tw=10^1$ (dotted), $10^2,10^3,10^4$ 
  and $\infty$ (full curve) obtained from the same equations. 
  }
\end{figure}

By combining the results derived in this section one obtains the 
FD plot and FDR presented in Fig.~\ref{fig:FDglobal}. Consistently 
with our expectations \cite{MayBerGarSol03,MayBerGarSol04,Pleimling04} 
a nontrivial limit FD plot exists for the global energy observable in the FA model. 
It defines the scaling form $\Xe(t,\tw) \sim \Xe(\tw/t)$ of the 
associated FDR. Since the energy-temperature step response $\chi_{\mathrm{e}}$ is negative,
the FD plot lies below the $\tilde{\chi} = 0$ axis. Because also the impulse response
$\rese$ is negative the slope of the plot, 
and thus the energy FDR $\Xe(t,\tw)$, are likewise {\em negative} throughout. 

As for the local defect observable it is crucial to use the correct 
parameterization of of the FD plot, with the observation time $t$ 
fixed and $\tw \in [0,t]$ the running parameter \cite{SolFieMay02, 
MayBerGarSol04}. Had we kept $\tw$ fixed and parameterized the 
plot with $t$, the non-monotonic shape of $\chi_{\mathrm{e}}(t,\tw) 
= \xse(t,\tw)$ shown in Fig.~\ref{fig:e} would have led to a non-monotonic 
FD plot, incorrectly suggesting a sign change in the FDR $\Xe(t,\tw)$. 
Only when parameterized with $\tw$ does the slope of the FD plot 
correspond to the FDR $X$. The introduction of no-field methods 
for the measurement of step response functions \cite{Chatelain03, 
Ricci03} has made it feasible to generate such FD plots with relative ease in numerical 
simulations. 

Visually the limit FD plot in Fig.~\ref{fig:FDglobal}
appears to be a straight line, but this in fact not the case. The slope 
becomes steeper from left to right and correspondingly the FDR 
$\Xe$ retains a dependence on the ratio $\tw/t$, c.f.\ 
Fig.~\ref{fig:FDglobal}. The latter is easily derived from the 
scaling of $\rse(t,\tw)$, Eq.~(\ref{equ:resesag}), and that of 
$\partial_{\tw} \core(t,\tw)$ from Eq.~(\ref{equ:coreag}), and 
has the explicit form 
\begin{equation}
\fl X_{\mathrm{e}}(u) = 
  - \left\{ \frac{2}{(1+u)^{3/2}} \left[ 1+\frac{2}{\pi} \arcsin u 
  \right] + \frac{2}{\pi \, u} \left[ \frac{(1-u)^{3/2}}{1+u} - 
  \frac{\arcsin\sqrt{u}}{\sqrt{u}} \right] \right\}^{-1}, 
  \label{equ:Xe}
\end{equation}
where $u = \tw/t$. The slope of the limit FD plot at the origin 
is given by $\Xe(\tw/t \to 1) = -(1+\sqrt{2}) \approx - 2.41$. 
This means that there is no quasi-equilibrium regime, a feature that seems to be 
generic for global observables, at least in 
coarsening systems \cite{MayBerGarSol03}. More importantly, 
the asymptotic energy FDR $\Xe^\infty$ is obtained by taking 
the opposite limit $\tw/t \to 0$, and this gives 
\begin{equation}
  \Xe^\infty = \lim_{u \to 0} X_{\mathrm{e}}(u) = -\frac{3 \, \pi}{6 \, \pi - 16} \approx -3.307. 
  \label{equ:Xeinfty}
\end{equation}
So indeed $\Xd^\infty = \Xe^\infty$: the asymptotic FDRs of local and
global observables are identical \cite{MayBerGarSol03,CalGam05,MayLeoBerGarSol06,
MayBerGarSol04,Pleimling04}. As for the case of local
defect observables our results will be exact to leading order for the
FA model in the long-time limit; this includes 
all aging expansions, the limit FD plots and the expressions for
$\Xe(u)$ and $\Xe^\infty$. Again, finite time data are included in Fig.~\ref{fig:FDglobal} 
only to give an indication of the speed of convergence 
(for free fermion DLPC) to the asymptotic scalings.~\footnote{We remark that 
the nonuniform convergence of $\Xe(t,\tw)$ in the right panel of 
Fig.~\ref{fig:FDglobal} is due to short-time contributions 
from $\Des(t,\tw)$ in Eq.~(\ref{equ:reses}) which yield $\rse(\tw,\tw) = 0$ 
and hence $\Xe(\tw,\tw) = 0$.}

While interesting in itself, the robustness of $X^\infty = \Xd^\infty = \Xe^\infty$ 
is also rather useful from a practical perspective: accurate measurement of 
$X^\infty$ in simulations based on the local defect observable is quite 
difficult as is clear from the FD plot in Fig.~\ref{fig:FDlocal}: the aging properties 
of $C_\mathrm{d}$ and $\chi_\mathrm{d}$ are disguised by quasi-equilibrium 
contributions. For the global energy observable, on the other hand, a nontrivial 
limit FD plot exists, see Fig.~\ref{fig:FDglobal}, and can be probed directly in 
simulations. Indeed, we have recently carried out such simulations for the FA model
and the results are in very good agreement with the theoretical predictions set out
above~\cite{MayLeoBerGarSol06}.

\section{Summary and Discussion}
\label{sec:summary}

In this paper we have analysed the nonequilibrium
fluctuation-dissipation (FD) behaviour of the
one-dimensional FA model after a quench to low temperature. The dynamics in
this regime is essentially a diffusion-coagulation process, and the
link from this to diffusion-annihilation and thence to the
Glauber-Ising chain allowed us to obtain exact scaling results for the
correlation and response functions of defects (mobile regions).  We
were thus able to resolve a puzzle posed by simulation results \cite{BuhGar02}: the
equilibrium behaviour suggested by numerical FD plots of local
quantities is only apparent and hides the underlying non-equilibrium effects. The
root cause of this is that the regime where the FDR $X$ differs
from unity becomes progressively compressed, as times grow, into a
corner of the plot. The FDR $X$ itself is nevertheless a nontrivial scaling
function of $\tw/t$. A second key observation was that local and
global defect observables give consistent values of the asymptotic FDR
$X^\infty$, reinforcing the generality of our findings for 
Ising models \cite{MayBerGarSol03,MayBerGarSol04,Pleimling04}.

An unexpected result of our analysis is that $X^\infty$ is, in fact,
negative. This seems counter-intuitive, certainly if one were to try to
interpret $T/X^\infty$ as an effective temperature. However, our analysis
showed that the negative value of $X^\infty$ can be traced to the
behaviour of the response functions, which are sensitive to the {\em
activated} nature of the dynamics \cite{DepSti05,JacBerGar06,MayLeoBerGarSol06}. 
In the global response function,
for example, the applied perturbation corresponds effectively to an
increase in temperature. In equilibrium this should {\em increase} the
energy. However, in the out-of-equilibrium dynamics the increased
temperature speeds up the relaxation of the energy and thus actually
{\em decreases} its value. In agreement with this intuition, we found
a negative (impulse) response $\rese(t,\tw)$. Except at short time
differences, it is $\tw$-independent and just given by the
time derivative (at time $t$) of the unperturbed energy relaxation:
the small ``extra time'' provided by the perturbation has the same
effect whenever it is applied.

As regards the wider applicability of our results, we have repeatedly
emphasised that the two-time correlation functions we have calculated
are exact for generic diffusion-limited pair coagulation (DLPC)
processes in one dimension, in the long-time scaling limit. In the
regime of small time differences, $\dt = t-\tw\ll\tw$, the behaviour is
easily rationalized in terms of non-interacting random walkers. Rather
less trivial is the behaviour in the opposite regime $t\gg\tw$. Here
we find scaling exponents identical to those for
diffusion-annihilation (DLPA) processes, but quantitatively different
scaling functions. The propagators used to derive these results are 
exact for the free fermion DLPC and DLPA processes. Higher order 
propagators could be obtained immediately from the results in \cite{MaySol04} 
and would form the key to analysing multi-particle and/or multi-time correlations 
in these processes for general initial conditions. 

Our response function results are similarly exact for generic DLPC
processes in the long-time limit. Here one has to bear in mind,
however, that the way the perturbation by the field enters can vary
between models. For example, a DLPC process can be realized as
the $q\to\infty$ limit of the one-dimensional Potts model \cite{DerZei96} with
Hamiltonian $\mathcal{H}=-\sum_i \delta_{s_i,s_{i+1}}$, $s_i\in\{1,\ldots,q\}$,
and Glauber rates. At low temperatures the domain walls
$n_i=1-\delta_{s_i,s_{i+1}}$ have DLPC dynamics with coagulation rate
1 and diffusion rate from bond $(i,i+1)$ to $(i+1,i+2)$ of
$1/2+(h_i+h_{i+1})/(4T)$, with the fields $h_i$ conjugate to $n_i$.
This is precisely the asymmetric part of the perturbation that we
found in the FA model; the symmetric part, which in the FA case
contains all the activation effects, is absent. Because the asymmetric
part of the response behaves similarly in DLPC and DLPA, see comments 
at the end of Sec.~\ref{sec:rad} and Sec.~\ref{sec:rae}, the
FD behaviour is then also like in the annihilation case \cite{MayBerGarSol03}, 
with an asymptotic FDR $X^\infty=0$.

Our results also shed light on a recently proposed method for
measuring response functions in spin systems \cite{LipCorZan05} 
and deriving associated nonequilibrium FD-relations.
This approach assumes a specific asymmetric form of the
perturbation, which in the FA case effectively amounts to discarding
the symmetric part of the response. As the discussion of the Potts
model above shows, this has drastic effects; in particular, it would
lead one to conclude incorrectly that the FA model has a vanishing
asymptotic FDR instead of the nontrivial value $X^\infty =
-3\pi/(6\pi-16)$. The method of \cite{LipCorZan05} and the
out-of-equilibrium FD-relation derived from it can therefore not be
accepted as generally valid.

It is worth noting that our result for the response function of
the FA model has recently been quoted as being consistent with the
predictions of a generalized form of local scale invariance \cite{HenEnsPle06}. 
This is indeed so
for the global response Eq.~(\ref{equ:resesag}) but {\em not} for the
local response function Eqs.~(\ref{equ:resdaag},\ref{equ:resdsag}), because
of the factors of $t+\tw$ that appear in both the symmetric and
asymmetric parts. This suggests to us that the agreement in the
global case is coincidental and cannot be taken to imply that
the FA model genuinely fits into the local scale invariance scheme.

In future work it will be interesting to extend our calculations to
the response and correlation of Fourier modes of the defect density,
which contain the global (energy) observable as the limiting case of
zero wavevector. Such Fourier decompositions have proved valuable in
the past in, e.g., Ising models \cite{MayBerGarSol03} and also
lend themselves naturally to a comparison with field-theoretical
results in two or more dimensions \cite{MayLeoBerGarSol06}. Taking a broader view, it
will be essential to establish how generic our finding of negative
response functions is: the link to activated dynamics suggests that,
once properly measured using, e.g., global observables, negative
responses may be much more common than previously thought. The
corresponding negative FDRs, while not having a natural
interpretation in terms of effective temperatures \cite{CugKurPel97}, 
may be worthwhile in and of themselves in elucidating the
nature of the activated dynamics.

\appendix

\section{Special Functions} 

Modified Bessel functions of integer order are most conveniently defined through 
the Fourier series of their generating function 
\begin{equation} 
  \rme^{t \cos(\cphi)} = \sum_{n=-\infty}^\infty I_n(t) \, \rme^{\rmi n \cphi} = 
    I_0(t) + 2 \sum_{n=1}^\infty I_n(t) \cos(n \cphi). 
  \label{equ:Ingen} 
\end{equation} 
Note that useful resummation properties follow immediately. For instance, setting 
$\cphi=0$ gives $\rme^t = I_0(t) + 2 \sum_{n=1}^\infty I_n(t)$. An explicit integral 
representation for the $I_n(t)$ is obtained by Fourier transformation, 
\begin{equation} 
  I_n(t) = \int_0^{2\pi} \frac{\rmd\cphi}{2\pi} \cos(n \cphi) \, \rme^{t \cos(\cphi)}. 
  \label{equ:Inint} 
\end{equation} 
Elementary properties of the $I_n(t)$ apparent from these equations are $I_{-n}(t) = I_n(t)$, 
$I_n(-t) = (-1)^n I_n(t)$ and $I_n(0) = \delta_{n,0}$. One shows that 
further 
\begin{eqnarray} 
  \partial_t  I_n(t) & = & \frac{1}{2} \left[ I_{n-1}(t) + I_{n+1}(t) \right], 
  \label{equ:Indiff} \\
  \frac{n}{t} I_n(t) & = & \frac{1}{2} \left[ I_{n-1}(t) - I_{n+1}(t) \right], 
  \label{equ:Inrec} 
\end{eqnarray} 
which we use extensively throughout the main text. From 
Eq.~(\ref{equ:Ingen}) we obtain the convolution property 
\begin{equation}
  I_m(t_1+t_2) = \sum_{n=-\infty}^\infty I_n(t_1) I_{m-n}(t_2). 
  \label{equ:Inconv} 
\end{equation}
The asymptotic behaviour of the modified Bessel functions of fixed and finite order $n$ 
in the limit $t \to \infty$ is given by \cite{Mathbook} 
\begin{equation}
  \rme^{-t} I_n(t) \sim \frac{1}{\sqrt{2 \pi t}}, 
  \label{equ:Inxlg} 
\end{equation}
whereas, if we simultaneously take $t,n \to \infty$ with $n^2/t$ fixed, 
\begin{equation}
  \rme^{-t} I_n(t) \sim \frac{1}{\sqrt{2 \pi t}} \, \rme^{-n^2/(2t)}. 
  \label{equ:Inxnlg} 
\end{equation}

Another important function in our analysis is $H_n(t)$, a full characterization of 
which can be found in \cite{MaySol04}; here we only summarize those features that are pertinent to 
the present context. First, $H_n(t)$ may be defined through 
\begin{equation}
  H_n(t) = \frac{1}{2} \int_0^t \rmd \tau \, \rme^{-\tau} \left[ I_{n-1}(\tau) - I_{n+1}(\tau) \right].
  \label{equ:Hndef}
\end{equation}
Clearly $H_{-n}(t) = -H_n(t)$ is odd in $n$ and hence $H_0(t) = 0$. Remarkably, all 
functions $H_n(t)$ can be decomposed into modified Bessel functions via the recursion 
\begin{eqnarray} 
	H_{n+1}(t) & = & H_n(t) + \delta_{n,0} - \rme^{-t} \left[ I_n(t) + I_{n+1}(t) \right], 
	\label{equ:Hnrec} 
\end{eqnarray} 
which applies for $n \geq 0$ and is derived by expressing $H_{n+1}(t) - H_n(t)$ 
via Eq.~(\ref{equ:Hndef}) and noting that the integrand becomes $\frac{1}{2} \, \rme^{-\tau} 
[ -I_{n-1}(\tau) + I_n(\tau) + I_{n+1}(\tau) - I_{n+2}(\tau)] = - \partial_\tau \, \rme^{-\tau} 
[I_n(\tau) + I_{n+1}(\tau)]$. Iterating Eq.~(\ref{equ:Hnrec}) we then have $H_1(t) = 1 - 
\rme^{-t} [I_0(t)+I_1(t)]$, $H_2(t) = 1 - \rme^{-t} [I_0(t)+2I_1(t)+I_2(t)]$, etc. As a 
consequence $H_n(t\to\infty) = 1$ for all $n \geq 1$. The latter is useful for deriving 
asymptotic expansions of $H_n(t)$: we rewrite Eq.~(\ref{equ:Hndef}) as $H_n(t) = 1 - 
\frac{1}{2} \int_t^\infty \rmd \tau \, \rme^{-\tau} \left[ I_{n-1}(\tau) - I_{n+1}(\tau) \right]$ 
so that $\tau \geq t$ in the integrand. Then, using Eqs.~(\ref{equ:Inrec}), (\ref{equ:Inxlg}) 
and (\ref{equ:Inxnlg}) we obtain that at fixed $n \geq 1$ and for $t \to \infty$ 
\begin{equation}
  H_n(t) \sim 1, 
  \label{equ:Hnxlg} 
\end{equation}
while for $t,n \to \infty$ with $n^2/t$ fixed, 
\begin{equation}
	H_n(t) \sim \Erfc \left( \frac{n}{\sqrt{2t}} \right) 
	\quad \mbox{where} \quad 
	\Erfc(z) = \frac{2}{\sqrt{\pi}} \int_z^\infty \rmd u \, \rme^{-u^2}. 
	\label{equ:Hnxnlg} 
\end{equation}
This defines the symbol $\Erfc(z) = 1 - \mathrm{erf}(z)$ for the complementary error function. 
It remains to add that beyond Eqs.~(\ref{equ:Hndef}) and (\ref{equ:Hnrec}) there is yet another 
useful representation for the function $H_n(t)$, viz. 
\begin{equation}
  H_{i_2-i_1}(2t) = 1 - \sum_{j_1 < j_2} \G{\ivec,\jvec}(t), 
  \label{equ:Hnsum} 
\end{equation}
which applies for $i_1 < i_2$. Consistently with the notation elsewhere the sum in Eq.~(\ref{equ:Hnsum}) 
runs over both $j_1,j_2 \in \mathbb{Z}$, but subject to the constraint $j_1 < j_2$. In order to establish the 
validity of Eq.~(\ref{equ:Hnsum}) we substitute the Green's function Eq.~(\ref{equ:G}), set $j_2 = j_1 + m$, shift 
$j = j_1-i_1$ and abbreviate $n = i_2-i_1 \geq 1$, which yields 
\begin{equation*}
  H_n(2t) = 1 - \rme^{-2t} \sum_{m=1}^\infty \sum_{j=-\infty}^\infty 
  \left[I_j(t) I_{j+m-n}(t) - I_{j-n}(t) I_{j+m}(t) \right]. 
\end{equation*}
After convolving the modified Bessel functions using Eq.~(\ref{equ:Inconv}) this expression reduces 
to $H_n(2t) = 1 - \rme^{-2t} \sum_{m=1}^\infty \left[ I_{m-n}(2t) - I_{m+n}(2t)\right]$. The remaining 
summation over $m \geq 1$ is telescopic and one easily verifies that it obeys the recursion 
Eq.~(\ref{equ:Hnrec}), thus proving the validity of Eq.~(\ref{equ:Hnsum}).

\section{Subdominant Responses} 
\label{sec:delta}

Here we establish subdominance of the corrections $\Delta_{\ivec,j}^{(a/s)}(\tw,\tw)$ 
to the instantaneous response functions for asymmetric and symmetric perturbations. 
These are given by 
\begin{eqnarray}
\fl \Delta_{\ivec,j}^{(a/s)}(\tw,\tw) = 
  \frac{1}{4} \left( \mp \delta_{j,i_1} + \delta_{j,i_1-1} \right) 
  \bra{e} \n_{i_1-1} \n_{i_1} E_{i_1+1,i_2} \, \rme^{\Wc \tw} \ket{1}
  \nonumber \\ 
  +\frac{1}{4} \left( + \delta_{j,i_2} \mp \delta_{j,i_2-1} \right) 
  \bra{e} E_{i_1,i_2-1} \n_{i_2-1} \n_{i_2} \, \rme^{\Wc \tw} \ket{1}. 
  \label{equ:Deltasa} 
\end{eqnarray}
The $-$ and $+$ alternatives for the signs correspond respectively to the asymmetric case, 
Eq.~(\ref{equ:Deltaa}), and the symmetric one, Eq.~(\ref{equ:Deltas}).
To evaluate Eq.~(\ref{equ:Deltasa}) we substitute $\n_i = 1 - E_{i,i+1}$ 
as usual, which reduces the averages to combinations of the form 
Eqs.~(\ref{equ:E1t},\ref{equ:E2t}). Then, making use of the recursion 
Eq.~(\ref{equ:Hnrec}) yields 
\begin{equation}
\fl \Da_{\ivec,j}(\tw,\tw) = - \frac{1}{4} A_{\ivec,j} \, \varphi_{i_2-i_1}(2\tw) 
  \quad \mbox{and} \quad 
  \Ds_{\ivec,j}(\tw,\tw) = \frac{1}{4} B_{\ivec,j} \, \varphi_{i_2-i_1}(2\tw), 
\end{equation}
with $A_{\ivec,j}$ and $B_{\ivec,j}$ the coefficients defined below 
Eq.~(\ref{equ:res1taD}) and (\ref{equ:res1tsD}), respectively, 
and 
\begin{equation}
\fl \varphi_n(2\tw) = \rme^{-2\tw} [I_0-I_2](2\tw) H_n(2\tw) + 
    \rme^{-4\tw} [(I_0+I_1)(I_{n-1}-I_{n+1})](2\tw). 
\end{equation}
At large $\tw$ and fixed $n^2/\tw$ this scales like 
\begin{equation}
  \varphi_n(2\tw) \sim \frac{1}{2\sqrt{\pi}\,\tw^{3/2}} \left[ 
  \Erfc\left(\frac{n}{2\sqrt{\tw}}\right) + \frac{2}{\sqrt{\pi}} 
  \frac{n}{2\sqrt{\tw}} \exp\left(-\frac{n^2}{4\tw}\right)\right], 
  \label{equ:phiscale} 
\end{equation}
which reduces correctly to the scaling $\varphi_n(2\tw) \sim 1/(2\sqrt{\pi}\,{\tw}^{3/2})$ 
for large $\tw$ but fixed $n$. Comparison to the leading contribution in the asymmetric 
response obtained from Eq.~(\ref{equ:res1taI}) shows that $\Da_{\ivec,j}(\tw,\tw)$ is 
uniformly -- with respect to $n = i_2 - i_1$ -- subdominant and may therefore be ignored. 
The situation is somewhat different for the symmetric case: at fixed $n$ and large $\tw$ 
the leading term in the symmetric response Eq.~(\ref{equ:res1tsI}) scales like 
$\mathcal{O}(n/{\tw}^{3/2})$ whereas 
$\Ds_{\ivec,j}(\tw,\tw) = \mathcal{O}(1/{\tw}^{3/2})$. For small $n \geq 1$ contributions from 
$\Ds_{\ivec,j}(\tw,\tw)$ cannot be neglected, however, the leading term in $\rs_{\ivec,j}(\tw,\tw)$ 
grows linearly with $n$ and thus quickly dominates $\Ds_{\ivec,j}(\tw,\tw)$. Specifically, once 
$n^2/\tw = \mathcal{O}(1)$ we have $\Ds_{\ivec,j}(\tw,\tw) = \mathcal{O}(1/{\tw}^{3/2})$ 
while the leading term in Eq.~(\ref{equ:res1tsI}) scales like $\mathcal{O}(1/\tw)$. Altogether 
contributions from $\Ds_{\ivec,j}(\tw,\tw)$ to the long time scaling of the symmetric 
instantaneous response become negligible when $n = i_2 - i_1 \gg 1$.

\end{document}